\documentclass[preprint]{aastex}
\usepackage{multirow}
\usepackage{graphicx}
\usepackage{color}
\usepackage{rotating}

\shortauthors{Wang}

\begin{document}
\title{Fundamental Performance of a Dispersed Fixed Delay Interferometer In Searching For Planets Around M Dwarfs}
\author{Ji Wang, Jian Ge, Peng Jiang, \& Bo Zhao}
\affil{Department of Astronomy, 211, Bryant Space Science Center, University of Florida, Gainesville,
    FL, 32611}
\email{jwang@astro.ufl.edu}

\begin{abstract}
We present a new method to calculate fundamental Doppler measurement limits 
 with a dispersed fixed-delay interferometer (DFDI) in the near infrared (NIR) wavelength region for 
searching for exoplanets around M dwarfs in the coming decade. It is based on 
calculating the $Q$ factor, a measure of flux-normalized Doppler
sensitivity in the fringing spectra created with DFDI. 
We calculate the $Q$ factor as a function of spectral resolution $R$, stellar projected rotational velocity $V \sin{i}$, stellar
effective temperature $T_{\rm{eff}}$  and optical path difference
(OPD) of the interferometer. We also compare the DFDI $Q$ factor to that for the
popular cross-dispersed echelle spectrograph method (the direct echelle (DE)  method). We find  that: 1, $Q_{\rm{DFDI}}$ is a factor of 1.5 to 3 higher than $Q_{\rm{DE}}$ at $R$ ranging from 5,000 to 20,000; 2, $Q_{\rm{DFDI}}$
and $Q_{\rm{DE}}$ converge at a very high $R$ ($R\geq$100,000); 3, $Q_{\rm{DFDI}}$ increases as $R$
increases and $V \sin{i}$ deceases; 4, for a given $R$, $Q_{\rm{DFDI}}$ increases as $T_{\rm{eff}}$ drops from 3100K to 2400K (M4V to M9V). 
 We also investigate how $Q_{\rm{DFDI}}$ is affected by
OPD and find that a 5 mm deviation from the optimal OPD does not significantly
affect $Q_{\rm{DFDI}}$ (10\% or less) for a wide range of $R$. 
Given the NIR Doppler measurement is likely to be detector-limited 
for a while, we introduce new merit functions, which is directly related to photon-limited RV
uncertainty, to evaluate Doppler performance with the DFDI and DE methods.
 We find that DFDI has strength in wavelength coverage and multi-object capability over the DE
for a limited detector resource. We simulate the performance
of the InfraRed Exoplanet Tracker (IRET) based on the DFDI design, being considered for the next generation
IR Doppler measurements. The
predicted photon-limited RV uncertainty suggests that IRET is
capable of detecting Earth-like exoplanets in habitable zone
around nearby bright M dwarfs if they exist.
 A  new method is developed to quantitatively estimate
 the influence of telluric lines on RV uncertainty. Our study shows that photon-limited RV uncertainty can be reached  if 
99\% of the strength of telluric lines can be removed from the measured 
stellar spectra. At low to moderate levels of telluric line strength
removal (50\% to 90\%), 
the optimal RV uncertainty is typically a factor of 2-3 times larger than photon-limited RV uncertainty.

\end{abstract}


\section{Introduction}

There are over 500 detected exoplanets as of June 2011 and about 80\% of
them were first discovered by the radial velocity (RV)
technique using ground-based Doppler instruments\footnotemark\footnotetext{http://exoplanet.eu/;
http://exoplanets.org/}. The popular Doppler instruments are based on the 
cross-dispersed 
echelle spectrogaph design, which we called the direct echelle (DE) method. In this method, 
the  RV signals are extracted by directly measuring the centroid
shift  of stellar absorption lines.  The fundamental photon-limited RV uncertainty using the DE method has been studied and reported by several 
research groups (e.g., ~\citet{Butler1996,Bouchy2001}).  
 The best achieved RV precision with the DE instrument is  less than 1 $\rm{m\cdot s}^{-1}$ RV precision 
with  bright and stable stars using the High Accuracy Radial Velocity Planet Searcher (HARPS) on the
European Southern Observatory 3.6 meter telescope~\citep{Bouchy2009}. 
 While DE is the most widely
adopted method in precision RV measurements, a totally different RV method using a dispersed fixed delay interferometer (DFDI) has also demonstrated its capability in discovering exoplanets~\citep{Ge2006,Fleming2010,Lee2011}. In this method, the RV signals are derived from phase shift of the interference
fringes created by passing stellar absorption spectra through a Michelson type interferometer with
fixed optical path difference (OPD) between the two interferometer arms~\citep{Erskine2000,Ge2002b,Ge2002,Erskine2003}. 
The stellar fringes are separated by  a
post-disperser, which is typically a medium-resolution spectrograph. 
 Doppler sensitivity of DFDI can be optimized by
carefully choosing the optical path difference of the
interferometer. The DFDI method is promising for its low cost,
compact size and potential for multi-object
capability~\citep{Ge2002}. ~\citet{VanEyken2010} discussed the theory
and application of DFDI in detail. However, its fundamental limit for Doppler measurements has not been well
studied before. In this paper, we introduce a method to calculate photon-noise limited Doppler measurement 
uncertainty in the near infrared (NIR) wavelength region, where we plan to apply the DFDI method for 
launching a Doppler planet survey around M dwarfs. 

IR Doppler planet surveys are very important to address planet characteristics around low mass stars, especially
M dwarfs. M dwarfs emit most of their photons in the NIR region. Due to lack of NIR Doppler techniques, 
only a few hundreds of the brightest M dwarfs have been searched for exoplanets using optical DE instruments~\citep{Wright2004,Endl2006,Zechmeister2009,Blake2010}.   
To date, only about 20 exoplanets around M dwarfs have been discovered compared to more than 500 exoplanets discovered around solar type
stars (i.e., FGK stars) despite of the fact that M dwarfs account for 70\% stars in local universe. Nonetheless, searching for planets around M
dwarfs is essential to answer questions such as the dependence of planetary properties on the spectral type of host stars. In addition, the smaller stellar mass of M dwarfs favors detection of rocky
planets in habitable zone (HZ) using the RV technique. However, the stellar absorption lines in NIR are not as sharp as those in the visible band. Recent study by ~\citet{Reiners2010} shows that precision RV measurements 
can only reach better Doppler precision in the NIR than in visible wavelength for M dwarfs with stellar
types later than M4. In this paper, we report results from our study on fundamental limits with 
the NIR Doppler technique using the DFDI method. 

This  paper is organized as follows. In \S \ref{sec:DFDI}, we briefly
review the principle of DFDI. In \S \ref{sec:SimulationMethod},
we describe the methodology of simulations in which we compute
photon-limited RV uncertainty for both DE and DFDI. In \S
\ref{sec:CompDeDFDI}, we compare the RV uncertainties of DE and DFDI under various conditions. In \S
\ref{sec:Performance}, we predict the photon-limited performance of a DFDI-based Doppler instrument in the NIR
and discuss its potential contributions to exoplanets search around M dwarfs. In
\S \ref{sec:Telluric}, we discuss the influence of telluric lines on
precision RV measurements and different methods of removing telluric
line effects. A summary of the study is given in \S \ref{sec:Conclusion}.

\section{Brief Review of DFDI}
\label{sec:DFDI}

\begin{figure}
\begin{center}
\includegraphics[width=8cm,height=10cm,angle=0]{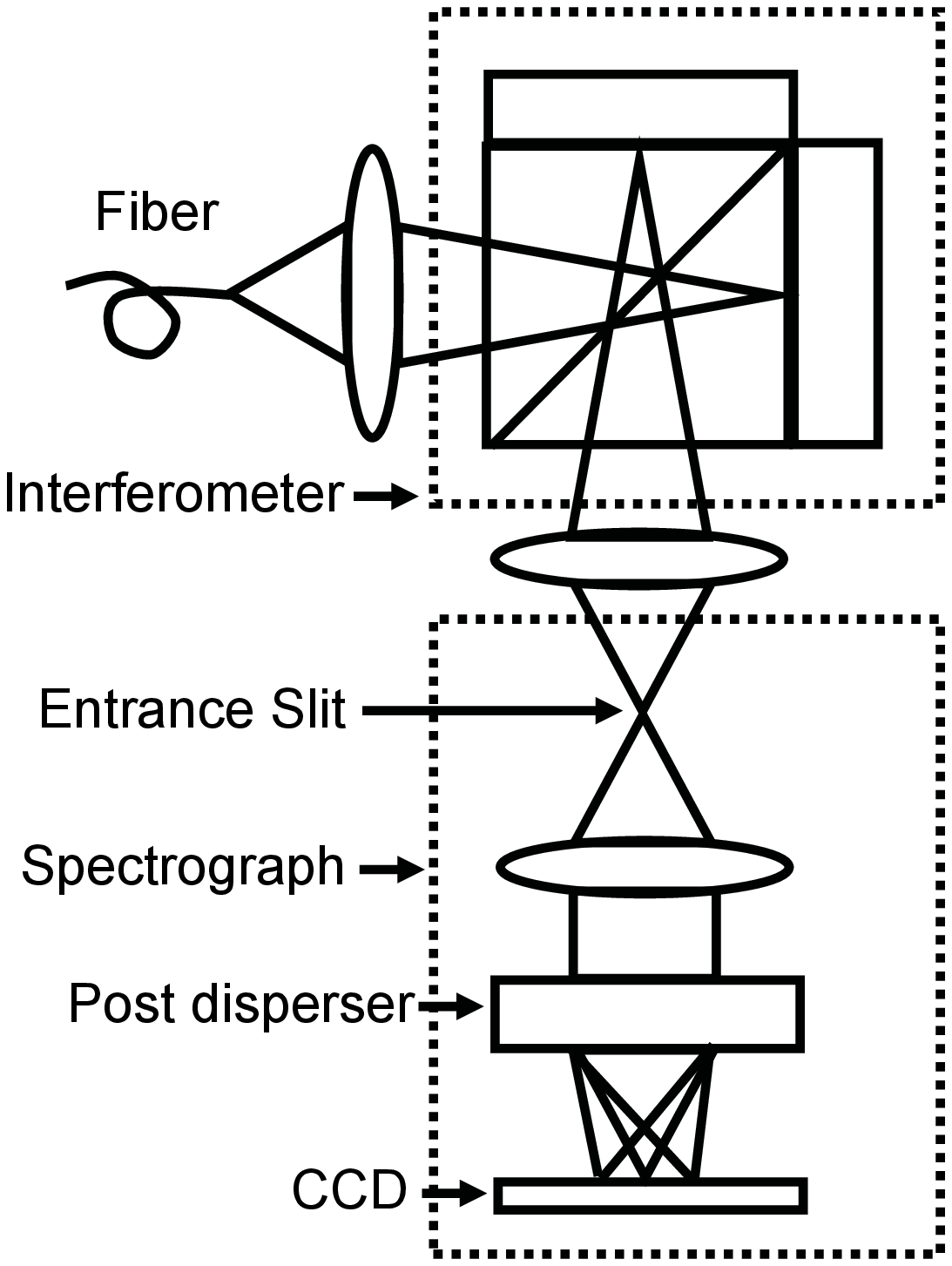}
\caption{A schematic layout of an RV instrument using the DFDI method.
\label{fig:DFDI_setup}}
\end{center}
\end{figure}

\begin{figure}
\begin{center}
\includegraphics[width=5cm,height=16cm,angle=270]{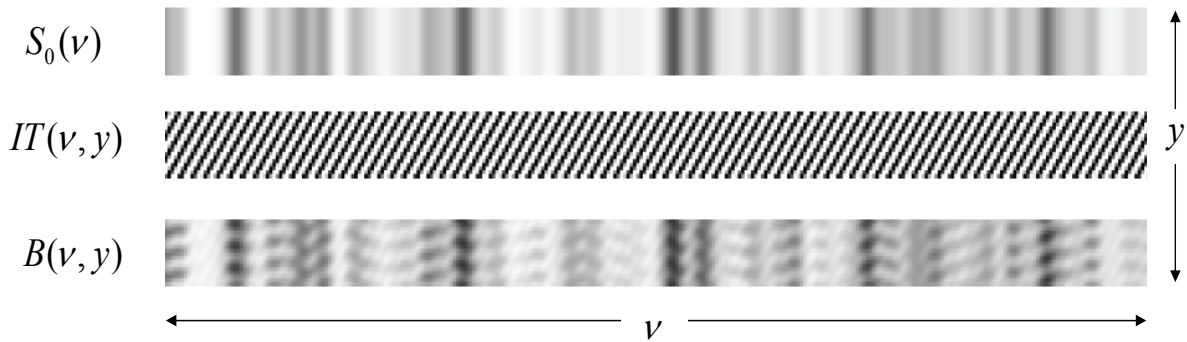}
\caption{DFDI Illustration. $S_0(\nu)$ is a stellar spectrum;
$IT(\nu,y)$ is interferometer transmission; $B(\nu,y)$ is the image
taken at a 2-D detector. $\nu$ is optical frequency and y is
coordinate of slit direction. DFDI measures RV by monitoring phase
shift of stellar absorption line fringes in the y direction (the slit direction). \label{fig:DFDI_illus}}
\end{center}
\end{figure}

The theory of DFDI has been discussed by several
papers~\citep{Ge2002, Erskine2003, VanEyken2010}. We briefly introduce the
principle of DFDI  in this section, readers may refer to previous
references for more detailed discussion. DFDI is realized by
coupling a fixed delay interferometer with a post-disperser (Fig.
\ref{fig:DFDI_setup}). The resulting fringing spectrum is recorded
on a 2-D detector. The formation of the final fringing spectrum is
illustrated in Fig. \ref{fig:DFDI_illus}. $B(\nu,y)$ is a
mathematical representation of the final image formed at the 2-D
detector and it is described by the following equation:
\begin{equation}
\label{eq:B_no_atm} B(\nu,y)=\bigg[\frac{S_0(\nu)}{h\nu}\times
IT(\nu,y)\bigg]\otimes LSF(\nu,R),
\end{equation}
where $S_0(\nu)$ is the intrinsic stellar spectrum and $\nu$ is
optical frequency. $S_0$ is divided by $h\nu$ to convert energy flux
into photon flux. $IT$ is the intensity transmission function
(Equation (\ref{eq:IT})), $y$ is the coordinate along the slit
direction which is transverse to dispersion direction, $\otimes$
represents convolution and $LSF$ is  the line spread function of the
post-disperser which is a function of $\nu$ and spectral resolution
$R$. In Equation (\ref{eq:IT}): $\gamma$ is visibility for a given frequency channel, the ratio of half of the peak-valley amplitude and the DC offset, which is determined by stellar flux $S_0(\nu)$; $c$ is the
speed of light; and $\tau$ is the optical path difference (OPD) of the
interferometer which is designed to be tilted along the slit direction
such that several fringes are formed along each $\nu$ channel (Middle, Fig. \ref{fig:DFDI_illus}). We
assume the LSF is a gaussian function (Equation (\ref{eq:LSF})),
$\Delta\nu=\nu/R/2.35$ because we assume that one resolution element
is equal to the FWHM of a spectral line.
\begin{equation}
\label{eq:IT}
IT=1+\gamma(\nu)\cdot\cos\bigg(\frac{2\pi\nu\tau(\nu,y)}{c}\bigg),
\end{equation}
\begin{equation}
\label{eq:LSF}
LSF(\nu_0,\Delta\nu)=\frac{1}{2\pi\nu^2}\exp\bigg(-\frac{(\nu-\nu_0)^2}{2\Delta\nu^2}\bigg).
\end{equation}

\begin{figure}
\begin{center}
\includegraphics[width=16cm,height=12cm]{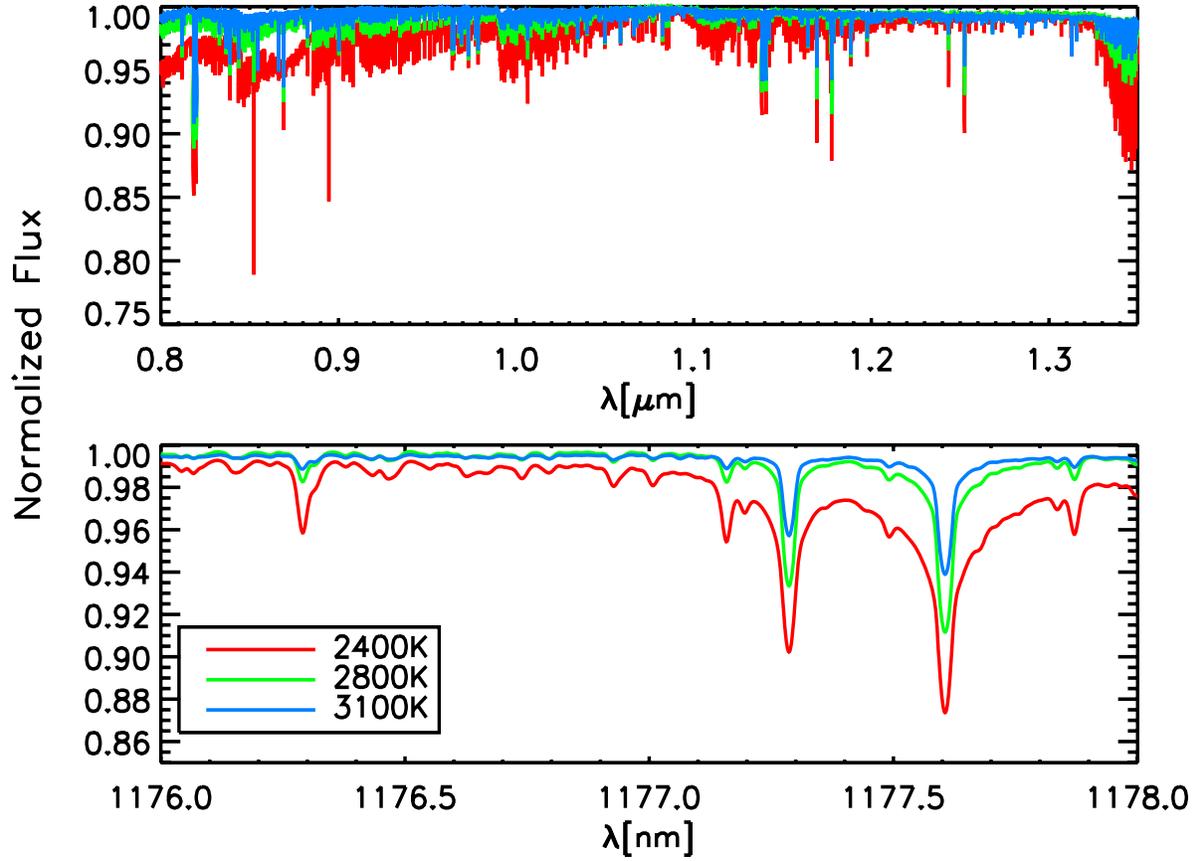}
\caption{Examples of synthetic M dwarf stellar spectra ($V
\sin{i}$=0 $\rm{km\cdot s}^{-1}$), which are generated by PHOENIX
~\citep{Hauschildt1999,Allard2001}. The top panel shows the spectra between 0.8-1.35 $\mu$m, the bottom panel shows an enlarged spectral region around 1177 nm showing stellar line profiles. \label{fig:Wav_Flux}}
\end{center}
\end{figure}

\begin{figure}
\begin{center}
\includegraphics[width=16cm,height=12cm]{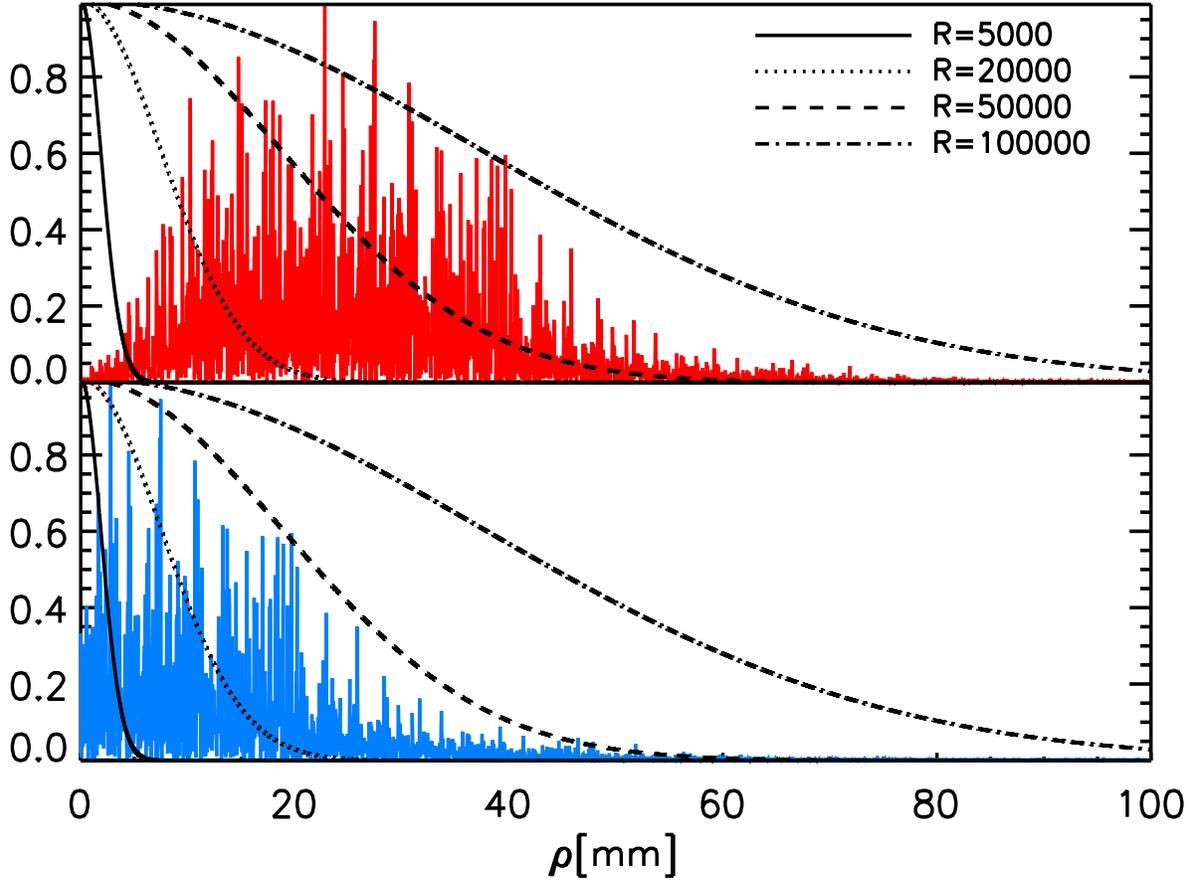}
\caption{Top: power spectrum of derivatives of stellar spectrum,
$\mathcal{F}[dS_0/d\nu]$, with $SRF(\rho)=\mathcal{F}[LSF(\nu,R)]$ for different $R$ overplotted, where LSF is line spread
function; Bottom: $\mathcal{F}[dS_0/d\nu]$ shifted by
$\Delta\rho=20mm$ using a fixed-delay Michelson interferometer. The $SRF$s for different $R$ are
 overplotted. \label{fig:Rho_Power_PSF}}
\end{center}
\end{figure}

Fig. \ref{fig:Wav_Flux} shows high-resolution (0.005 \AA\ spacing)
synthetic spectra of M dwarfs with solar metallicity~\citep{Hauschildt1999,Allard2001}. $T_{\rm{eff}}$ ranges from 2400K to 3100K, and $\log g$ is 4.5. No rotational
broadening is added in the spectrum. Most absorption lines are shallow with
FWHMs of several tenths of an \AA. Since RV information is
embedded in the slope of an absorption line,  sharp and
deep lines contain more RV information than broad and shallow lines.
Mathematically, the slope is the derivative of flux as a function of
optical frequency, i.e., $dS_0/d\nu$. The power spectrum of $dS_0/d\nu$ is obtained by Fourier transform. According to properties of Fourier
transform, $\mathcal{F}[dS_0/d\nu]=(i\rho)\cdot \mathcal{F}[S_0]$,
where $\mathcal{F}$ manifests Fourier transform, $i$ is the unit of
imaginary number and $\rho$ is the representation of $\nu/c$ in
Fourier space. We plot $\mathcal{F}([dS_0/d\nu]$ in Fig.
\ref{fig:Rho_Power_PSF} as well as the spectral response function
(SRF), which is $\mathcal{F}[LSF]$. SRF at $R=5,000$ drops drastically toward high
spatial frequency (high $\rho$ value) such that it misses most of the RV
information contained in stellar spectrum. As $R$ increases, SRF
gradually increases toward high $\rho$ where the bulk of RV
information is stored. A spectrograph with $R$ of $\sim$100,000 is
capable of nearly completely extracting RV information. Unlike DE, DFDI can
shift $\mathcal{F}[dS_0/d\nu]$ by an amount determined by the OPD of
the interferometer~\citep{Erskine2003}. For example, Fig. \ref{fig:Rho_Power_PSF} also shows the power spectrum of $\mathcal{F}([dS_0/d\nu]$ of a fringing 
spectrum obtained with a DFDI instrument with a 20 mm optical delay, which shifts 
$\mathcal{F}[dS_0/\nu]$ by 20 mm. In this case, RV information has been shifted from the original 
high spatial frequencies to low spatial frequencies which can
 be resolved by a spectrograph with a low or medium $R$ in
DFDI.

\section{Simulation Methodology}
\label{sec:SimulationMethod}

In the DE method, an  efficient way based on a spectral quality factor ($Q$) was introduced by~\citet{Bouchy2001} to calculate the fundamental uncertainty in the  Doppler 
measurements.  The $Q$ factor 
is a measure of spectral profile information within a given wavelength 
region considered for Doppler measurements. 
Here we develop a similar method  to calculate $Q$ values for
the DFDI method. Instead of representing the spetral line profile information in the DE method,
 the $Q$ factor in our DFDI method represents stellar fringe profile information. 
We use high resolution (0.005 \AA\ spacing) synthetic stellar
spectra generated by PHOENIX code\citep{Hauschildt1999,Allard2001}
because observed spectra of low mass stars do not have high enough
resolution and broad effective temperature coverage.
\citet{Reiners2010} have conducted several comparisons between
synthetic spectra generated by PHOENIX and the observed spectra. They concluded that the synthetic spectra are accurate enough for RV measurement uncertainty calculation. We used
synthetic stellar spectra of solar abundance with $T_{\rm{eff}}$ ranging from
2400K to 3100K (corresponding spectral type from M9V to M4V) and a surface gravity $\log g$ of 4.5. The $Q$ factor is calculated for a series of 10 nm spectral slices from
800 nm to 1350 nm. We artificially broaden spectra with $V
\sin{i}$ from 0 $\rm{km\cdot s}^{-1}$ to 10 $\rm{km\cdot s}^{-1}$ assuming a limb darkening index of
0.6, which is a typical value for an M dwarf. We convolve the
rotational broadening profile with each spectral slice of 10 nm
to obtain a rotationally-broadened spectrum. We assume a
Gaussian LSF which is determined by
spectral resolution $R$ (Equation (\ref{eq:LSF})). After
artificial rotational broadening and LSF convolution, we rebin
each spectral slice according to 4.2 pixels per resolution element (according to the optical design of IRET by~\citet{Zhao2010})
to generate the final 2D image on a detector  based on which we compute
the $Q$ factor.

\subsection{Photon-limited RV Uncertainty of DE}
\label{sec:echelle}

~\citet{Bouchy2001} described a method of calculating the $Q$ factor
for the DE method. We briefly introduce the method here and the reader can
refer to ~\citet{Bouchy2001} for more details. Let $S_0(\nu)$
designate an intrinsic stellar spectrum. $A_0$, a
digitalized and calibrated spectrum, is considered as a noise-free template
spectrum for differential RV measurement, which is related to
$S_0(\nu)$ via the following equation:
\begin{equation}
A_0(i)=\frac{S_0(\nu)}{h\nu}\otimes LSF(\nu),
\end{equation}
where $i$ is pixel number and $S_0$ is divided
by $h\nu$ to convert energy flux into photon flux. Another
spectrum $A$ is taken at a different time with a tiny Doppler shift,
which is small relative to the typical line width of an intrinsic
stellar absorption. Assuming that the two spectra have the same
continuum level, Doppler shift is given by:
\begin{equation}
\label{eq:Doppler}
\frac{\delta v}{c}=\frac{\delta\nu}{\nu},
\end{equation}
where $c$ is speed of light and $\nu$ is optical frequency. The
overall RV uncertainty for the entire spectral range is given by~\citep{Bouchy2001}:
\begin{eqnarray}
\label{eq:overall_Doppler} \frac{\delta v_{rms}}{c} & = &
{Q}^{-1}\cdot\bigg[{{\displaystyle\sum_i A_0(i)}}\bigg]^{-1/2}=\frac{1}{Q\sqrt{N_{e^-}}},
\end{eqnarray}
where $Q$ is defined as:
\begin{equation}
\label{eq:q_factor} Q\equiv\Bigg[\frac{{\displaystyle\sum_i
W(i)}}{{\displaystyle\sum_i A_0(i)}}\Bigg]^{1/2},
\end{equation}
and $W(i)$ is expressed as:
\begin{equation}
W(i)=\frac{\bigg(\frac{\partial
A_0(i)}{\partial\nu(i)}\bigg)^2\nu(i)^2}{A(i)}.
\end{equation}
The $Q$ factor is independent of photon flux and represents
extractable Doppler information given an intrinsic stellar spectrum and
instrument spectral resolution $R$. According to Equation
(\ref{eq:overall_Doppler}), we can calculate photon-limited RV
uncertainty given the $Q$ factor and photon flux $N_{e^-}=\displaystyle\sum_i
A_0(i)$ within the spectral range.

\subsection{Photon-limited RV Uncertainty of DFDI}
\label{sec:edi}

A new method of calculating the $Q$ factor for DFDI is developed and discussed
here. After a digitalization process, a 2-D flux distribution expressed by Equation
(\ref{eq:B_no_atm}) is recorded on a 2-D detector in DFDI. The digitalization process involves distributing photon flux into each pixel according to: 1) pixels per resolution element (RE);
2) spectral resolution; 3) number of fringes along slit. $B_0(i,j)$, which is a noise-free template, is then calculated. $B(i,j)$ is a frame taken at a different time with a tiny Doppler shift. $i$ is the pixel number along the dispersion
direction, and $j$ is the pixel number along the slit direction. The observable
intensity change at a given pixel $(i,j)$ in DFDI is expressed by:
\begin{eqnarray}
\label{eq:after_disperser} B(i,j)-B_0(i,j) & = & \frac{\partial
B_0(i,j)}{\partial\nu(i)}\delta\nu(i) \nonumber \\ & = &
\frac{\partial B_0(i,j)}{\partial\nu(i)}\cdot\frac{\delta
v}{c}\cdot\nu(i).
\end{eqnarray}
The Doppler shift is measured by monitoring the intensity
change at a given pixel in the equation:
\begin{equation}
\label{eq:after_disperser} \frac{\delta
v}{c}=\frac{B(i,j)-B_0(i,j)}{\frac{\partial
B_0(i,j)}{\partial\nu(i)}\cdot\nu(i)}.
\end{equation}
Frame $B_0$ is assumed to be a noise-free template and the noise of
frame $B$ is the quadratic sum of the photon noise and the detector
noise $\sigma_D$:
\begin{equation}
\label{eq:noise_2d}
B_{rms}(i,j)=\sqrt{B(i,j)+\sigma^2_D}.
\end{equation}
Equation (\ref{eq:noise_2d}) is approximated under photon-limited
conditions as $B_{rms}(i,j)=\sqrt{B(i,j)}$. Therefore, the RV
uncertainty at pixel $(i,j)$ is given by:
\begin{equation}
\label{eq:Doppler}
\frac{\delta
v_{rms}(i,j)}{c}=\frac{\sqrt{B(i,j)}}{\frac{\partial
B_0(i,j)}{\partial\nu(i)}\cdot\nu(i)}.
\end{equation}
The overall RV uncertainty for the entire spectral range is given
by:
\begin{eqnarray}
\label{eq:overall_Doppler_2d} \frac{\delta v_{rms}}{c} & = & \bigg[{{\displaystyle\sum_{i,j} \big(\frac{\delta
v_{rms}(i,j)}{c}\big)^{-2}}}\bigg]^{-1/2} \nonumber \\
 & \equiv & \bigg[{{\displaystyle\sum_{i,j}
W(i,j)}}\bigg]^{-1/2}\nonumber \\ & \equiv &
{Q}^{-1}\cdot\bigg[{{\displaystyle\sum_{i,j} B_0(i,j)}}\bigg]^{-1/2} \nonumber \\
 & = & \frac{1}{Q\sqrt{N_{e^-}}},
\end{eqnarray}
where
\begin{equation}
W(i,j)\equiv\frac{\bigg(\frac{\partial
B_0(i,j)}{\partial\nu(i)}\bigg)^2\nu(i)^2}{B(i,j)},
\end{equation}
and
\begin{equation}
\label{eq:q_factor_2d}
Q\equiv\Bigg[\frac{{\displaystyle\sum_{i,j}
W(i,j)}}{{\displaystyle\sum_{i,j} B_0(i,j)}}\Bigg]^{1/2}.
\end{equation}
Equation (\ref{eq:q_factor_2d}) calculates the $Q$ factor for the DFDI
method, which is also independent of flux and represents the Doppler
information that can be extracted with the DFDI method. According to
Equation (\ref{eq:overall_Doppler_2d}), we can calculate
photon-limited RV uncertainty given the $Q$ factor and photon flux
$N_{e^-}=\displaystyle\sum_{i,j} B_0(i,j)$ within the spectral range.

\section{Comparison between DE and Optimized DFDI}
\label{sec:CompDeDFDI}
\subsection{Optimized DFDI}
\label{sec:OptiDFDI}

\begin{figure}
\begin{center}
\includegraphics[width=16cm,height=12cm]{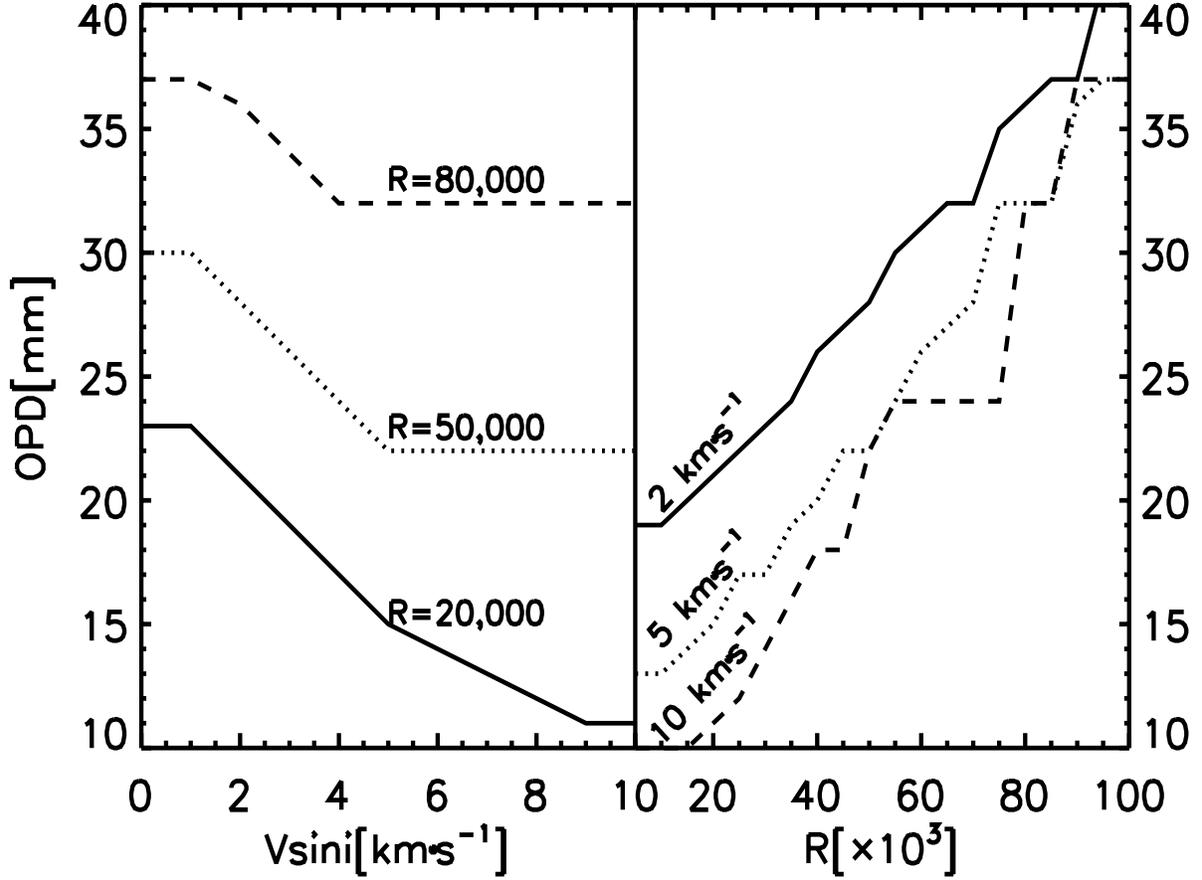} \caption{Optimal OPD correlation with $V\sin i$ (left) and spectral resolution $R$ (right). Optimal OPD for $R$=20,000 (solid), 50,000 (dotted), 80,000 (dashed) are used on the left panel. $V\sin i$=2 (solid), 5 (dotted), 10 (dashed) $\rm{km}\cdot\rm{s}^{-1}$ are assumed on the right panel. Complete results of optimal OPD can be found in Table \ref{tab:OPD_choise}. $T_{\rm{eff}}$ influence on optimal OPD is not significant, $T_{\rm{eff}}$=2800 K is adopted in the plot. \label{fig:OPD_Vsini_R}}
\end{center}
\end{figure}

Optical Path Difference (OPD) of a fixed delay interferometer is a
crucial parameter that affects the Doppler sensitivity of a DFDI
instrument (Ge 2002). An optimized OPD can help increase the instrument
Doppler sensitivity. We calculate the optimal OPD for spectra of various
$T_{\rm{eff}}$ and $V \sin{i}$ at different spectral resolutions (Table \ref{tab:OPD_choise}). We assume a wavelength range from 800 nm to 1350 nm and an OPD range from 10mm to 41mm with a step size of 1mm in the calculation as
described in \S \ref{sec:edi}. Optimal OPD is selected as the one
which results in the highest $Q$ factor value.  Increasing $V
\sin{i}$ or decreasing $R$ naturally broadens absorption lines, decreasing 
the coherence length of each stellar absorption line (Ge 2002).  Consequently, our  simulations show in general
that the optimal OPD decreases with increasing $V \sin{i}$ or decreasing $R$
values (Fig. \ref{fig:OPD_Vsini_R}). We also note that $T_{\rm{eff}}$ influence on optimal OPD is not significant.

\begin{figure}
\begin{center}
\includegraphics[width=16cm,height=12cm]{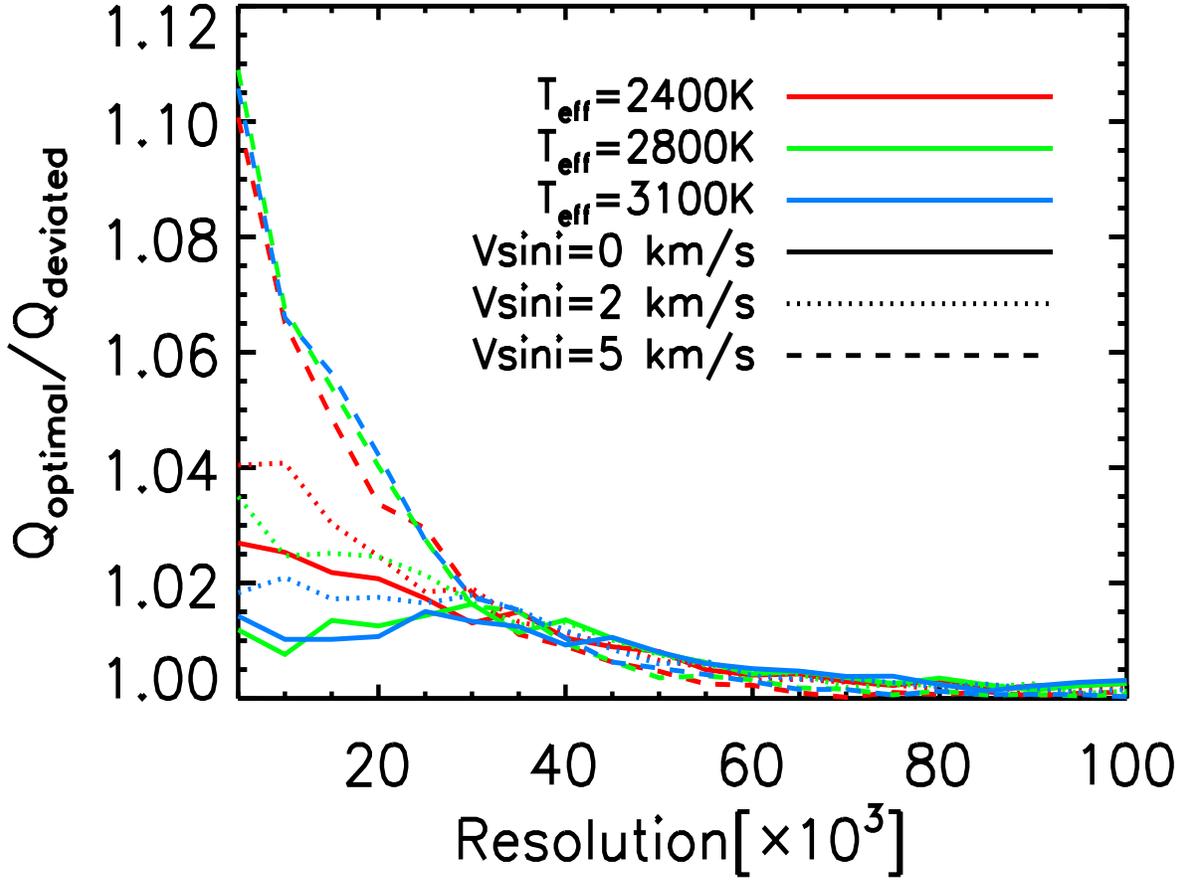}
\caption{Ratio of $Q_{\rm{optimal}}$ and $Q_{\rm{deviated}}$ as a function of
spectral resolution, where $Q_{\rm{optimal}}$ is $Q$ factor at optimal OPD
and $Q_{\rm{deviated}}$ is $Q$ factor when OPD is deviated from
$Q_{\rm{optimal}}$ by 5 mm. Different line styles represent different $V\sin i$ while colors indicate different $T_{\rm{eff}}$. \label{fig:OpdGain_Res}}
\end{center}
\end{figure}

We investigate how the $Q$ factor is affected if OPD is deviated from
the optimal value. We calculate the $Q$ factor when the actual OPD is
deviated from the optimal OPD by 5mm.  We
 choose the lower value of the two $Q$s from the 5 mm 
deviation from the optimal delay (both positive and negative sides) as
$Q_{\rm{deviated}}$. We plot the ratio of $Q_{\rm{optimal}}$ and
$Q_{\rm{deviated}}$ as a function of spectral resolution $R$ in Fig.
\ref{fig:OpdGain_Res}. We found that deviating OPD by 5mm does not
result in severe degradation of the $Q$ factor. The maximum
degradation is 1.115 and occurs at $R$ of 5,000 and $V \sin{i}$ of 5 $\rm{km\cdot s}^{-1}$ for a star with $T_{\rm{eff}}$ of 2800 K.
The degradation can be compensated by increasing the integration time by 24\% ($1.115^2$) to reach the same
photon-limited Doppler
precision according to equation (13). The degradation becomes smaller as $R$ increases. As shown in Fig. \ref{fig:Rho_Power_PSF}, 
DFDI shifts the power spectrum of $dS_0/d\nu$ by an amount
determined by the interferometer OPD so that the SRF has a reasonable
response at a region where most of the RV information is
contained. The SRF broadens as $R$ increases.
Therefore, it can still recover most of the RV information in
a stellar spectrum even if OPD is deviated from the optimal value.
At low and medium resolutions (5,000 to 20,000), $Q$ factor degradation
becomes larger as $V \sin{i}$ increases. This is because 
rotational broadening removes the high frequency signal from
a stellar spectrum, which makes the region containing most of the RV
information more sensitive to the choice of OPD as the power spectrum distribution becomes narrower due loss of high $\rho$ component.

\subsection{Influence of Spectral Resolution $R$}
\label{sec:DFDI_Res}

\begin{figure}
\begin{center}
\includegraphics[width=16cm,height=12cm]{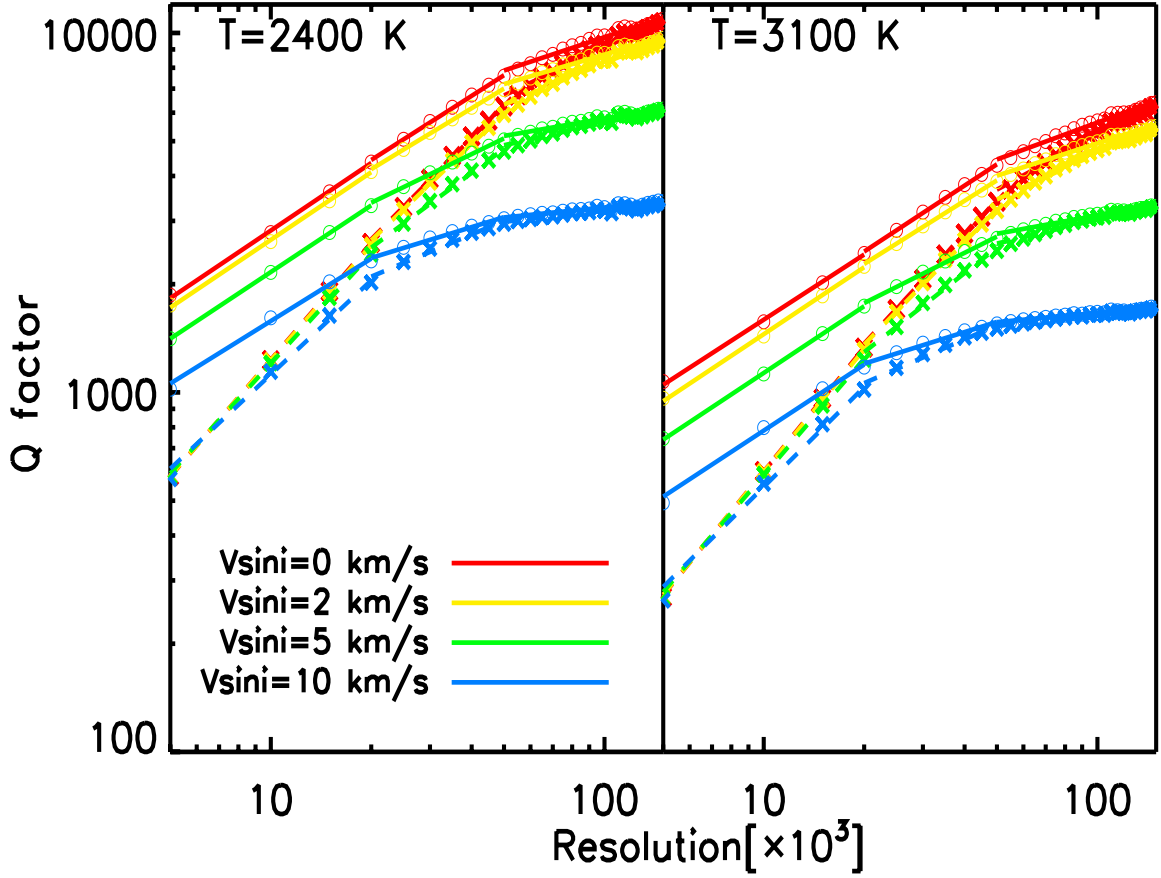} \caption{$Q$ factor as a function of spectral resolution.
(left: $T_{\rm{eff}}=2400K$; right: $T_{\rm{eff}}=3100K$. Open circles represent $Q_{\rm{DFDI}}$; crosses
represent $Q_{\rm{DE}}$; solid lines are best power-law fits for $Q_{\rm{DFDI}}$;
dashed lines are best power-law fits for $Q_{\rm{DE}}$)\label{fig:Q_Res}}
\end{center}
\end{figure}

In theory, a spectrograph with an infinitely high resolution would be able to
extract all the RV information contained in a stellar spectrum. However, in pratice,
 it is impossible to completely recover the RV information with a spectrograph with a finite spectral resolution 
whose  spectral response function
 drops at the high spatial frequency end. Although the power spectrum of the derivative of the
stellar spectrum is shifted to the low frequency region where most of
the RV information is carried, the power spectrum is still broad in the spatial
frequency ($\rho$) domain (see Fig. \ref{fig:Rho_Power_PSF}). Thus, high $R$ can help to extract more RV information. 
In a wavelength coverage from 800 nm to 1350 nm, we calculate $Q$
values for stellar spectra with $V \sin{i}$ of 0 ,2 ,5 and 10 $\rm{km\cdot s}^{-1}$ at different $R$ (5,000 to 150,000 with a step of 5,000)
in order to investigate the dependence of $Q$ on $R$ (Fig.
\ref{fig:Q_Res}). We find that more RV
information (higher $Q$ factor) can be extracted as $R$ increases. $Q$ factors for DFDI and
DE converge at high $R$ because the spectral response function is
wide enough in the $\rho$ domain to cover the region rich in RV
information, not affected by the power spectrum shifting involved in DFDI. In
addition, the $Q$ factor at a given $R$ increases as $T_{\rm{eff}}$ drops
from 3100K to 2400K, which is largely due to stronger molecular absorption
features in the I, Y and J bands (see Fig. \ref{fig:Wav_Flux}).

We divide $R$ into three regions, low resolution (5,000 to
20,000), medium resolution (20,000 to 50,000) and high resolution (50,000 to
150,000). We use a power law to fit $Q$ for both DFDI and DE as a
function of $R$. The power indices $\chi$ of three regions for $T_{\rm{eff}}=2400K$ are presented in Table \ref{tab:PowerLawR}. At low $R$ region, $\chi$
remains roughly a constant for 0 $\rm{km\cdot s}^{-1}$ $\leq V \sin i\leq$ 5 $\rm{km\cdot s}^{-1}$,
but it drops for stars with $V \sin i$ of 10 $\rm{km\cdot s}^{-1}$ indicating stellar
absorption lines begin to be resolved even at low $R$. At higher $R$
regions, $\chi$ decreases as $V \sin i$ increases, a reduced value
of $\chi$ implies diminishing benefit brought by increasing $R$. Stellar absorption lines are broadened by stellar
rotation, and they are resolved at a certain $R$ beyond which increasing
$R$ does not significantly gain Doppler sensitivity. Overall,
$\chi$ for DE is larger than that of DFDI, especially for low and medium $R$. In
other words, $Q_{\rm{DFDI}}$ is less sensitive to a change of $R$, and the DFDI
instrument can extract relatively more Doppler information 
 at low or medium spectral resolution than the DE method. For example, for slow
rotators ($V \sin i$=2 $\rm{km\cdot s}^{-1}$) at the low $R$ region (R=5,000-20,000),
$Q_{\rm{DFDI}}\propto R^{0.63}$. Doppler sensitivity $\delta v_{rms}$ is
inversely proportional to two factors: $Q$ and $\sqrt{N_{e^-}}$
according to Equation (\ref{eq:overall_Doppler}) and
(\ref{eq:overall_Doppler_2d}), where $N_{e^-}$ is the total photon count collected by 
 the CCD detector. $N_{e^-}\propto(S/N)^2\cdot N_{\rm{pixel}}$, where
$S/N$ is the average signal to noise ratio per pixel, and $N_{\rm{pixel}}$ is
total number of pixels. Note that $N_{e^{-}}\propto R$ if the 
wavelength coverage, S/N per pixel and the resolution sampling are fixed.
Therefore, $\delta v_{rms}\propto R^{-0.63-0.5}=R^{-1.13}$ for DFDI. In comparison, $\delta v_{rms}\propto R^{-1.57}$ for DE given the same wavelength coverage and S/N per pixel. The power law is
consistent with the previous theoretical work by ~\citet{Ge2002} and
~\citet{Erskine2003}. 

\begin{figure}
\begin{center}
\includegraphics[width=16cm,height=12cm]{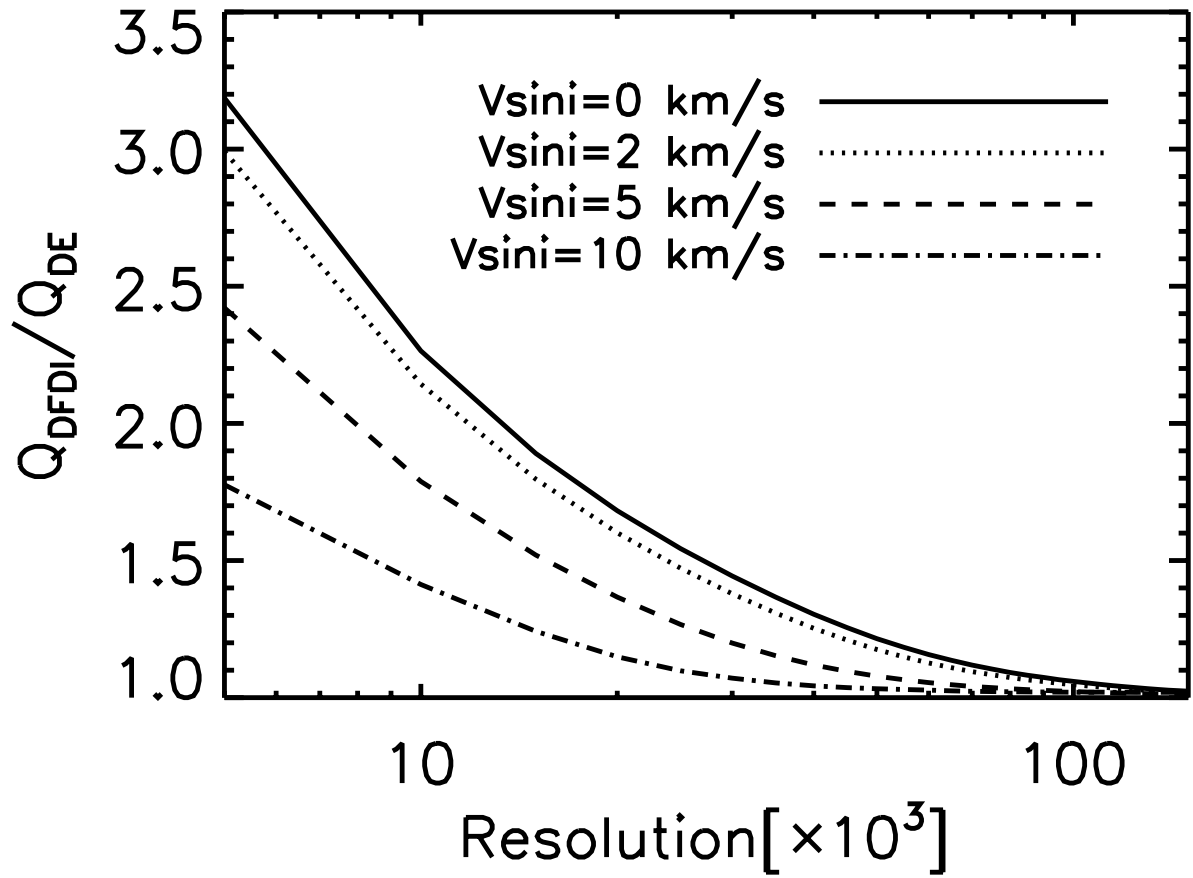} \caption{Improvement of $Q_{\rm{DFDI}}$ over $Q_{\rm{DE}}$ as a function of
spectral resolution. \label{fig:Qgain_Res}}
\end{center}
\end{figure}

We compare $Q$ factors for both DFDI and DE at given $R$ values, and
the results are shown in Fig. \ref{fig:Qgain_Res}. For very slow
rotators ($0\ \rm{km\cdot s}^{-1}\le V \sin{i}\le2\ \rm{km\cdot s}^{-1}$), the advantage of DFDI
over DE is obvious at low and medium $R$ (5,000 to 20,000) because the
center of the power spectrum of the derivative of the stellar spectrum is at a high frequency domain which 
cannot be covered in DE due to the limited frequency response range of its SRF
at low and medium $R$. The improvement of DFDI is $\sim$3.1 times
($R$=5,000), $\sim$2.4 times ($R$=10,000) and $\sim$1.7 times ($R$=20,000) respectively.
In other words, optimized DFDI with $R$ of 5,000, 10,000 and 20,000 is
equivalent to DE with $R$ of 16,000, 24,000 and
34,000 respectively in terms of Doppler sensitivity for the same wavelength coverage, S/N per pixel and 
spectral sampling (otherwise, see more discussions in \S \ref{sec:CCD_size}, the gain with the DFDI would be more 
significant for a fixed detector size and exposure time).
Overall, DE with the same spectral resolution as DFDI at R=5,000-20,000 requires $\sim$3-9 times longer exposure time  
 to reach the same Doppler sensitivity as DFDI if both instruments have the same wavelength coverage and same
detection efficiency (i.e., $N_{e^-}$ is the same). 
 The improvement of DFDI at $R$=20,000-50,000
is not as noticeable as at the low $R$ range. The
difference between DFDI and DE becomes negligible when $R$ is over
100,000. In other words, the advantage of the DFDI over DE 
gradually disappears as $R$ reaches high resolution domain 
($R>50,000$). In addition,  the improvement for relatively faster rotators ($5\ \rm{km\cdot s}^{-1}\le V \sin{i}\le10\
\rm{km\cdot s}^{-1}$) with DFDI is less significant than it is for
slow rotators.

\subsection{Influence of Detector Pixel Numbers}
\label{sec:CCD_size}

In the NIR, detector pixel number is typically smaller than the optical detector. Furthermore, the total cost 
for an NIR array is much higher than an optical detector with the same pixel number. In the foreseeable future, detector size may be one of the major limitations for Doppler sensitivity improvement. We study
the impact of the limited detector resource on the Doppler measurement sensitivity. Using the same detector resource, we find that
it is fair to compare their Doppler performance
for the same target with the same exposure to understand strength and weakness for each method 
although DFDI and DE are totally different Doppler techniques.
According to Equation (\ref{eq:overall_Doppler}) and
(\ref{eq:overall_Doppler_2d}), we define a new merit function,  
\begin{equation}
\label{eq:Qprime1} Q^\prime=Q\cdot\sqrt{N_{e^-}},
\end{equation}
to study photon-limited Doppler performance for both methods with the same
detector size. Note that the newly defined merit function is directly related to photon-limited RV uncertainty, i.e., inverse proportionality. $N_{e^-}$ is calculated by Equation (\ref{eq:photon_flux}):
\begin{equation}
\label{eq:photon_flux} N_{e^-}=\frac{F_*\cdot\eta\cdot S_{\rm{tel}}\cdot
t_{\rm{exp}}}{2.512^{m_J}},
\end{equation}
in which $F_*$ is the photon flux in the wavelength coverage region
$\Delta\lambda$ of an $m_J=0$ star with the unit of $\rm{photons}\cdot s^{-1}\cdot cm^{-2}$;
$\eta$ is instrument total throughput; $S_{\rm{tel}}$ is the effective surface area of the
telescope; $t_{\rm{exp}}$ is the time of exposure; and $m_J$ is the J
band magnitude. 

\begin{figure}
\begin{center}
\includegraphics[width=16cm,height=12cm]{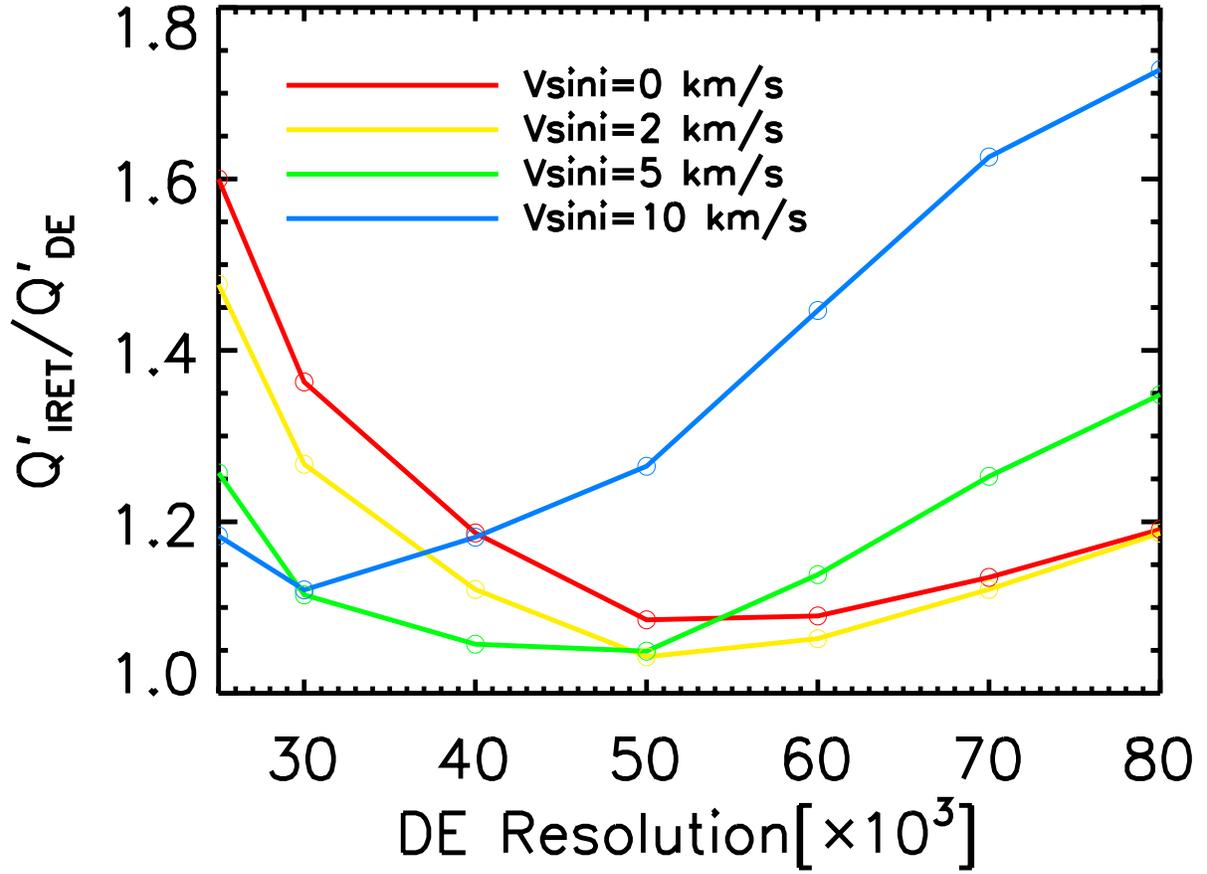} \caption{Comparison of $Q^\prime_{\rm{IRET}}$ and $Q^\prime_{\rm{DE}}$
at different $R$. Note that $Q^\prime=Q\cdot\sqrt{N_{e^-}}$.  Different color represents different rotational velocity. \label{fig:Q_IRET_Q_DE_Res}}
\end{center}
\end{figure}

Here we use IRET as an example to illustrate strengths of the DFDI method for Doppler measurements. 
IRET adopts the DFDI method and has a wavelength
coverage from 800 nm to 1350 nm and a spectral
resolution of 22,000. For a fixed detector size (i.e., total number of detector pixels) and fixed number of
pixels to sample each resolution element, the total wavelength coverage of a Doppler instrument, $\Delta\lambda$, is
\begin{equation}
\label{eq:bandpass}
\Delta\lambda=\frac{N_{\rm{pix}}}{P_{\rm{order}}}\cdot{\frac{\lambda_c}{R\cdot {N_S}}},
\end{equation}
where $N_{\rm{pix}}$ is total number of pixels available on a CCD detector and $N_S$ is the number of pixels per resolution element, $\lambda_c$ is the central wavelength and $P_{\rm{order}}$ is the number of pixels sampling each pixel channel between spectral orders\footnote{In principle, each 
frequency channel for both DFDI and DE instruments can be designed identical. In practice, the DFDI instrument tends to use $\sim$ 20 pixels to sample fringes in the slit direction, which is only for measurement convenience, not a requirement.  In fact, ~\citet{Muirhead2010} has demonstrated a phase-stepping method which does not require sample fringes in the slit direction.)}. Equation
(\ref{eq:bandpass}) shows that $\Delta\lambda$ is inversely
proportional to $R$. Table \ref{tab:R_DeltaLambda}
gives the relation of $R$ and $\Delta\lambda$ assuming $N_{\rm{pixel}}$, $P_{\rm{order}}$ and $N_S$ as constants. $\lambda_c$ is
set to be 1000 nm because it is approximately the center of the Y band. We
calculated the ratio of $Q'_{\rm{DFDI}}$ and $Q'_{\rm{DE}}$ in which we
use the photon flux of a star with a $T_{\rm{eff}}$ of 2400K (Fig.
\ref{fig:Q_IRET_Q_DE_Res}). $Q^\prime_{\rm{IRET}}$ is consistently higher
than $Q^\prime_{\rm{DE}}$ regardless of $R$ of the DE instrument. In
other words, IRET is able to achieve lower photon-limited RV
uncertainty compared to a DE instrument with the same detector.
The result seems to be different from the conclusion we drew in \S
\ref{sec:DFDI_Res}, in which we compare $Q$ factors of the same $R$ and
$\Delta\lambda$ and reached a conclusion
that DFDI with an $R$ of 22,000 is equivalent to DE with an $R$ of
35,000 (a factor of 1.6 gain) for the same wavelength coverage and the same detection efficiency (Fig.
\ref{fig:Q_Res}). The key difference between this case and the earlier case is the fixed detector 
resource instead of fixed total collected photon numbers. Since lower spectral resolution allows
to cover more wavelengths, more photons will be collected for the same instrument detection efficiency 
for both DFDI and DEM.  Note that $Q^\prime$ consists of two components, $Q$ and $N_{e^-}$. For a given number of pixels on the detector, $N_{e^-,\rm{DFDI}}$ is higher than $N_{e^-,\rm{DE}}$ due
to the larger wavelength coverage. In addition,
$Q_{\rm{DFDI}}(\Delta\lambda_{\rm{DFDI}})$ is more than $Q_{\rm{DE}}(\Delta\lambda_{\rm{DE}})$. Consequently, we see in Fig.
\ref{fig:Q_IRET_Q_DE_Res} that $Q^\prime_{\rm{IRET}}$ is higher than
$Q^\prime_{\rm{DE}}$ at all $R$ of a DE instrument.  Fig. \ref{fig:Q_IRET_Q_DE_Res} also shows that 
the minimum of $Q^\prime_{\rm{DFDI}}/Q^\prime_{\rm{DE}}$ is dependent of
 $V\sin i$. The ratio of $Q^\prime$ reaches a minimum
($Q^\prime_{\rm{DE}}$ reaches a maximum) around an $R$ of 50,000 for slow
rotators ($V \sin{i}\le5$ $\rm{km\cdot s}^{-1}$). It increases at the low $R$ end
because the spectrograph has not yet resolved stellar absorption
lines.  On the other hand, the ratio increases at the high $R$ end because of fewer
photons (see Table \ref{tab:R_DeltaLambda}). For fast rotators ($V\sin i$=10 $\rm{km\cdot s}^{-1}$), the
ratio reaches a minimum around $R$ of 30,000.

For a slow rotator ($V \sin i$=2 $\rm{km\cdot s}^{-1}$) at low $R$ region (R=5,000-20,000), $Q_{\rm{DFDI}}\propto R^{0.63}$. Since Doppler sensitivity $\delta v_{rms}$ is inversely proportional to two factors: $Q$ and $\sqrt{N_{e^-}}$
according to Equation (\ref{eq:overall_Doppler}) and
(\ref{eq:overall_Doppler_2d}), the Doppler sensitivity becomes nearly independent of spectral resolution for the 
DFDI method ($\propto R^{-0.13}$) if the detection size (or total number of pixels) is fixed. This indicates that we can use quite moderate resolution spectrograph to disperse the stellar fringes produced by the interferometer in a DFDI instrument while maintaining high Doppler sensitivity. This opens a major door for multi-object Doppler measurements using the DFDI method as proposed by Ge (2002). 
In comparison, the Doppler sensitivity for the DE method still strongly depends on spectral resolution for a fixed number of detector pixels ($\propto R^{-0.57}$), 
indicating  that higher spectral resolution will offer better Doppler sensivity.

\subsection{Influence of Multi-Object Observations}
\label{sec:multi_obj}
As discussed in \S \ref{sec:DFDI_Res} and \ref{sec:CCD_size}, the DFDI instrument can be 
designed to have a moderate resolution spectrograph coupled with a Michelson
type interferometer. Moderate spectral resolution allows a single order 
spectrum or a few order spectra to cover a broad wavelength region in the 
NIR region while keeping the Doppler sensivitiy similar to 
a high resolution DE design which requires a large detector array to cover 
spectra from a single target. This indicates that the DFDI method has much greater 
potential for accommodating multiple targets on the same detector as proposed by 
Ge (2002) than the DE instrument. 
In order to evaluate the potential impact of
multi-object DFDI instruments,  we redefine the merit function $Q^{\prime\prime}$ as:
\begin{equation}
\label{eq:Qprime2} Q^{{\prime\prime}}=Q\cdot\sqrt{N_{e^-}}\cdot
N_{obj}^{\alpha},
\end{equation}
where $N_{obj}$ is the number of objects that can be monitored
simultaneously, and $\alpha$ is the index of importance for
multi-object observations.  From
the perspective of photon count and S/N, multi-object observations are
equivalent to an increase of $N_{e^-}$, and thus $\alpha$ is 0.5. However, from an
observational efficiency point of view, $Q^{\prime\prime}$ should be
proportional to $N_{obj}$ because the more objects are observed
simultaneously, the quicker the survey is accomplished, and $\alpha$ is therefore 
equal to 1. 

\begin{figure}
\begin{center}
\includegraphics[width=16cm,height=12cm]{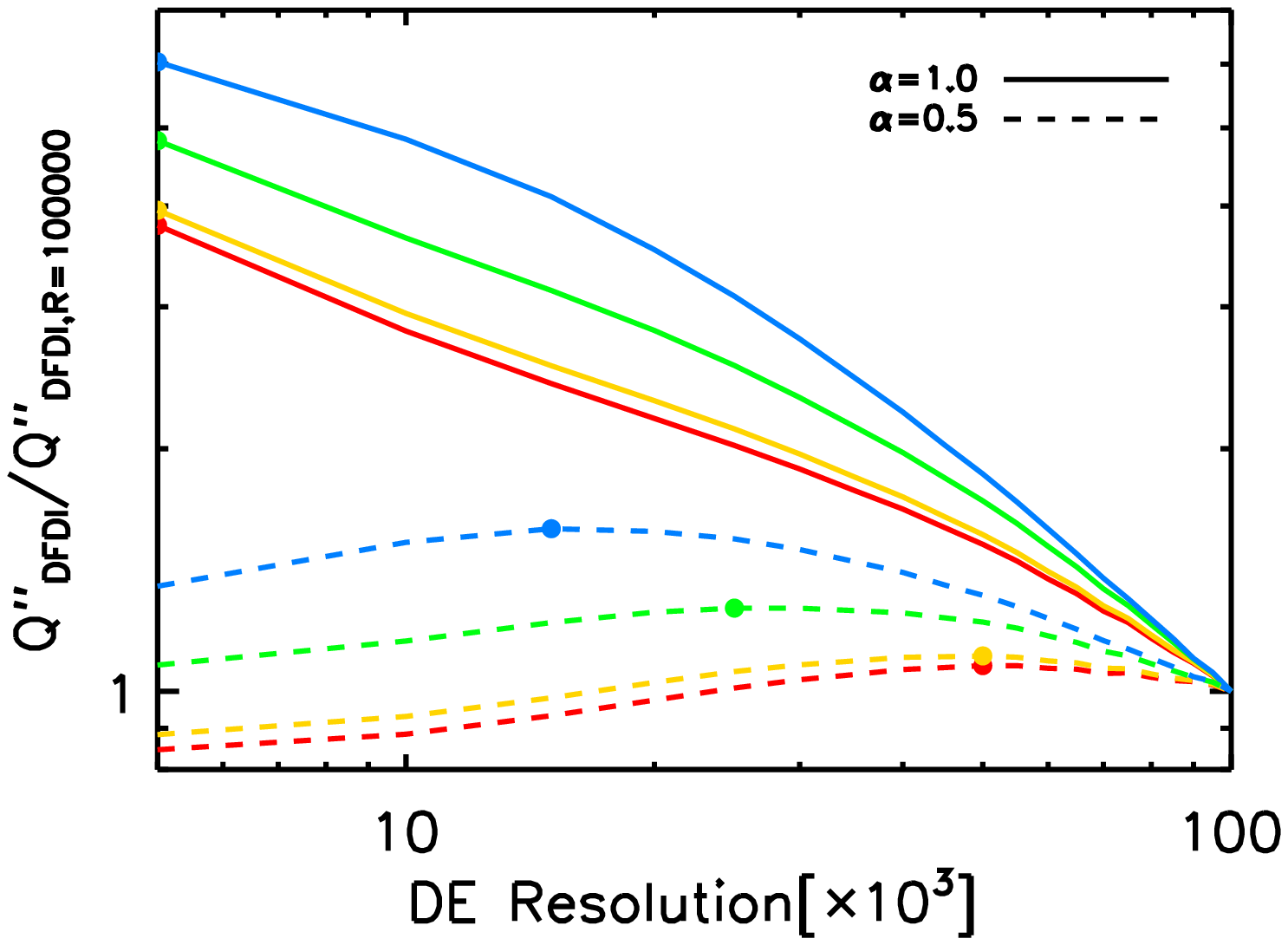} \caption{Comparison of $Q^{\prime\prime}_{\rm{DFDI}}$ and $Q^{\prime\prime}_{\rm{DFDI, R=100,000}}$ at different $R$. Note that $Q^{\prime\prime}=Q\cdot\sqrt{N_{e^-}}\cdot
N_{obj}^{\alpha}$. The maximum of each curve is indicated by filled circle. Different color represents different rotational velocity, the same as Fig. \ref{fig:Q_IRET_Q_DE_Res}. \label{fig:Q_IRET_Q_DE_Res_multi}}
\end{center}
\end{figure}

\begin{figure}
\begin{center}
\includegraphics[width=16cm,height=12cm]{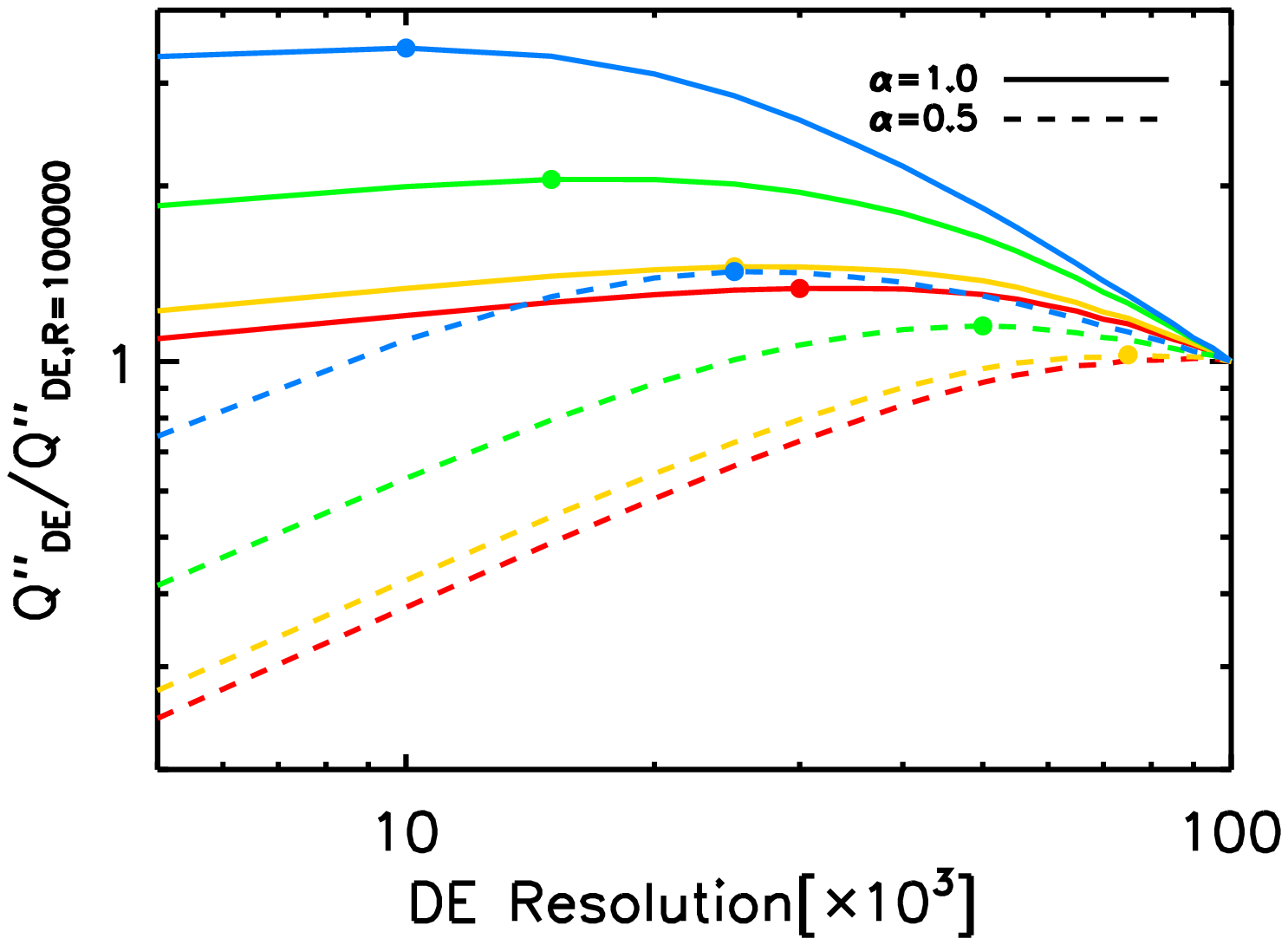} \caption{Comparison of $Q^{\prime\prime}_{\rm{DE}}$ and $Q^{\prime\prime}_{\rm{DE, R=100,000}}$ at different $R$. Note that $Q^{\prime\prime}=Q\cdot\sqrt{N_{e^-}}\cdot
N_{obj}^{\alpha}$. The maximum of each curve is indicated by filled circle. Different color represents different rotational velocity, the same as Fig. \ref{fig:Q_IRET_Q_DE_Res}. \label{fig:Q_IRET_Q_DE_Res_multi_2}}
\end{center}
\end{figure}

We assume a detector that covers from 800 nm to 1350 nm at $R$=100,000 so that we can use the $Q$ factors obtained in \S \ref{sec:DFDI_Res}.  $N_{e^-}$ is a constant since we assume identical $\Delta\lambda$. $N_{obj}$ is inversely proportional to the number of pixels per object which is proportional to spectral resolution $R$ (Equation (\ref{eq:bandpass})). Note that we do not require $N_{obj}$ to be an integer because we can, in principle, 
fit a fraction of spectrum on a detector to make full use of the detector. $Q^{\prime\prime}$s for both DFDI and DE are calculated. Fig. \ref{fig:Q_IRET_Q_DE_Res_multi} shows the ratio of  $Q^{\prime\prime}$ and $Q^{\prime\prime}_{R=100,000}$ for DFDI under two different assumptions of $\alpha$. For $\alpha$=0.5, i.e., increase of $N_{obj}$ is equivalent to photon gain, only a slight improvement is achieved if the detector is used for multi-object observations at lower resolution than 100,000. In comparison, from a survey efficiency point of view (i.e., $\alpha$=1),  we see a factor of $\sim$4-6 times boost of $Q^{\prime\prime}$ in multi-object observations. The truncation at $R$=5,000 is due to a practical reason that a lower resolution than 5,000 is rarely used in planet survey using RV techniques. On the other hand, similar calculation is also conducted for the DE method (Fig. \ref{fig:Q_IRET_Q_DE_Res_multi_2}), in which we find that high resolution single object observation is an optimal operation mode for DE from a perspective of photon gain ($\alpha$=0.5). At $\alpha$=1, the increase of $Q^{\prime\prime}$ is a factor of $\sim$3 at the most. 

We compare the maximum of $Q^{\prime\prime}$ for both DFDI and DE at different $V\sin i$ in Table \ref{tab:QPrimePrime}. At $\alpha$=0.5, the advantage of DFDI over DE is $\sim$1.1 for a wide range of $V\sin i$. In other words, from the photon gain point of view, there is no significant difference between DFDI and DE in multi-object RV instruments. However, from the survey efficiency point of view ($\alpha$=1),
 we see a factor of 3 boost of $Q^{\prime\prime}$ in DFDI for slow rotators ($V\sin i\leq$2$\rm{km}\cdot\rm{s}^{-1}$), suggesting 9 times faster in terms of survey speed. For fast rotators (i.e., $V\sin i$=10$\rm{km}\cdot\rm{s}^{-1}$), the boost drops to 1.78. Our study confirms that the 
DFDI method has an advantage for multi-object RV measurements over the DE method as suggested by ~\citet{Ge2002}.

\subsection{Influence of Projected Rotational Velocity $V \sin{i}$}
\label{sec:DFDI_Vsini}

\begin{figure}
\begin{center}
\includegraphics[width=16cm,height=12cm]{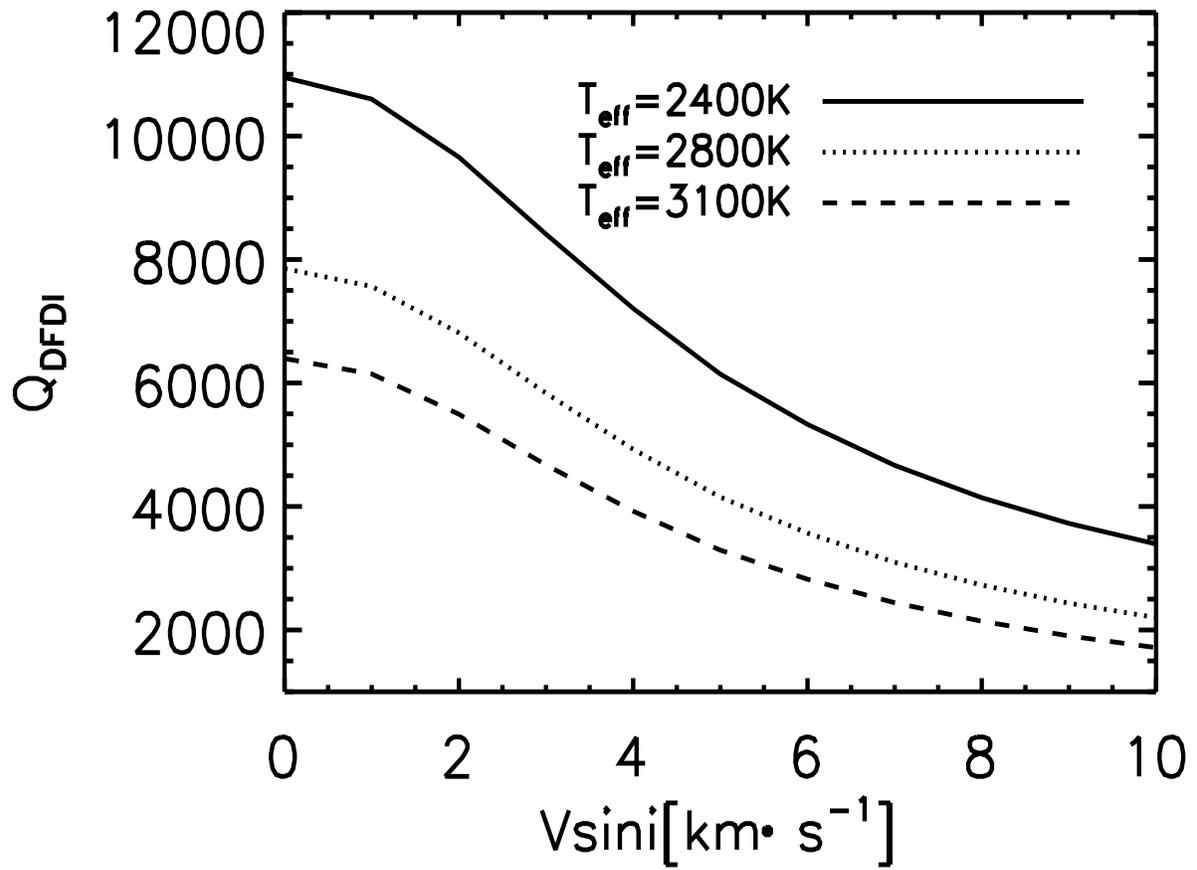} \caption{$Q_{\rm{DFDI}}$ as a function of $V \sin{i}$.
\label{fig:Q_Vsini}}
\end{center}
\end{figure}

Projected rotational velocity $V \sin{i}$ broadens stellar
absorption lines and thus reduces the $Q$ factor. We carry out
simulations calculating $Q$ factors of different $V \sin{i}$ 
(0 $\rm{km\cdot s}^{-1}\leq V \sin{i}\le$10 $\rm{km\cdot s}^{-1}$) 
at $R$=150,000 and various $T_{\rm{eff}}$. We assume a wavelength range from 800 nm to 1350 nm.
The results are shown in Fig. \ref{fig:Q_Vsini}. The $Q$ factor
decreases as $V \sin{i}$ increases. It is clear that  slow rotators would
be better targets to reach  higher
photon-limited RV precision because the spectrum of a slow rotator
contains more Doppler information.

\section{IRET Performance and its Survey Capability}
\label{sec:Performance}

Here we use the IRET as an example to demonstrate the Doppler sensitivity with the DFDI method and its capability for an NIR survey for exoplanets around M dwarfs. IRET has a spectral
resolution $R$ of 22,000. It works mainly in the I, Y and J bands
from 800 to 1350 nm. 4.2 pixels sample one resolution element
and 25 pixels sample the fringes in the slit (y) direction~\citep{Zhao2010}. 

\begin{figure}
\begin{center}
\includegraphics[width=16cm,height=12cm]{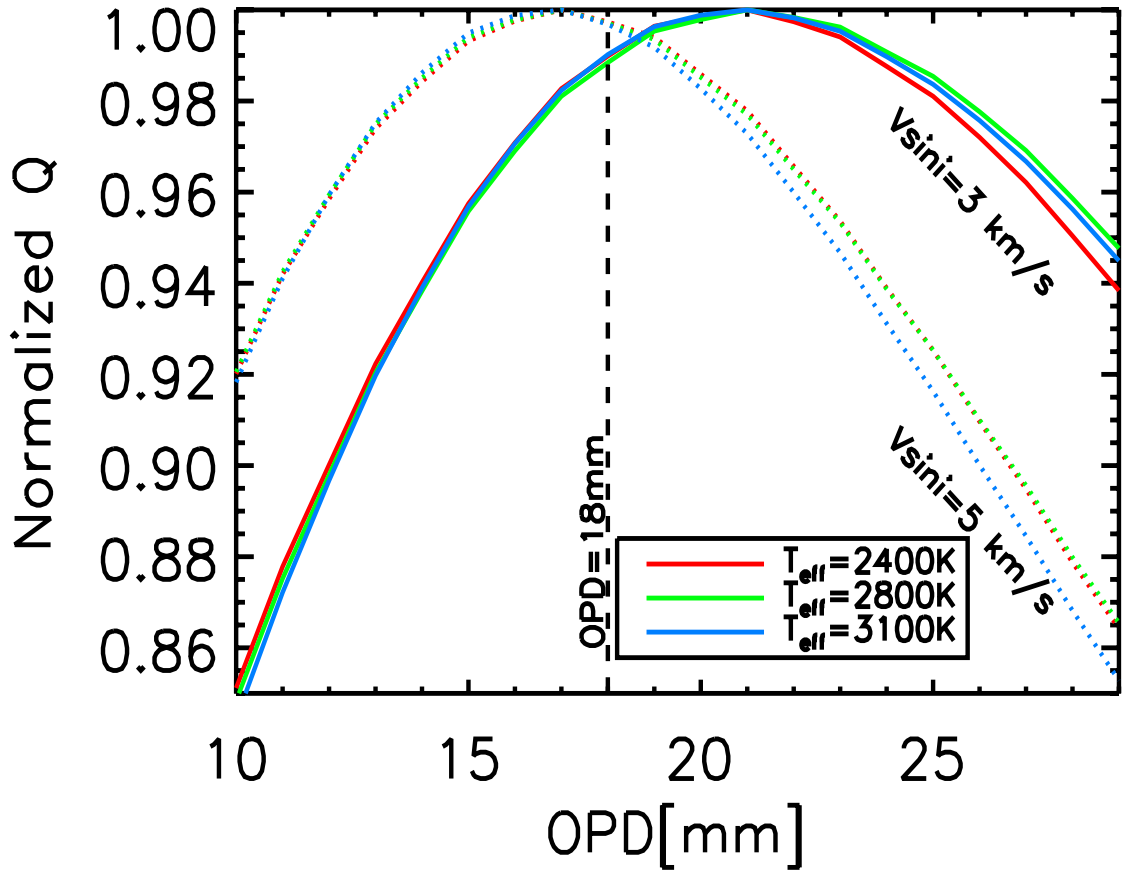}
\caption{OPD optimization for IRET. OPD is optimized at $R=22,000$,
$V \sin{i}$ between 3 $\rm{km\cdot s}^{-1}$ and 5 $\rm{km\cdot s}^{-1}$. \label{fig:Q_temp}}
\end{center}
\end{figure}

\begin{figure}
\begin{center}
\includegraphics[width=16cm,height=12cm]{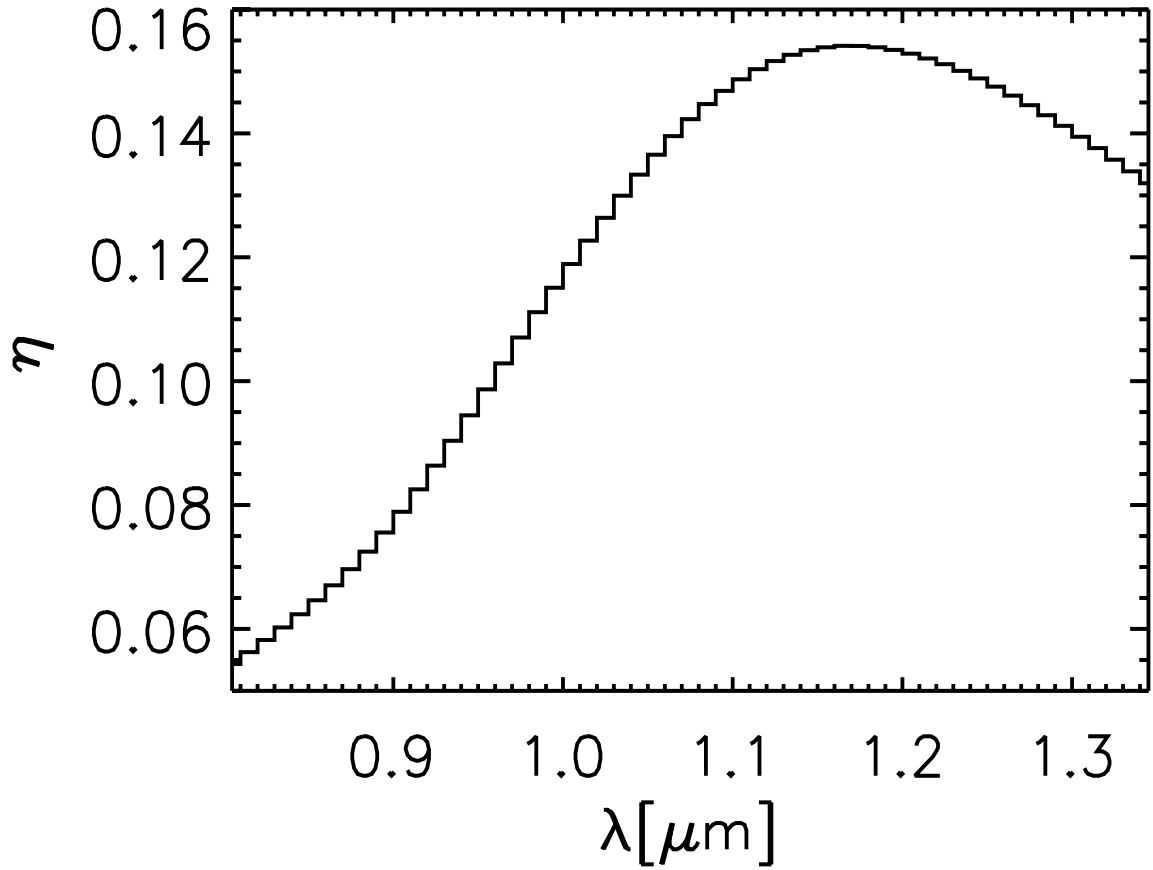} \caption{Instrument throughput $\eta$ as a function of wavelength
(800 to 1350 nm). The telescope transmission is 80\%. It does not include
atmospheric tranmission, fiber coupling, transmission and seeing losses.  \label{fig:Wav_Eta}}
\end{center}
\end{figure}

\begin{figure}
\begin{center}
\includegraphics[width=16cm,height=12cm]{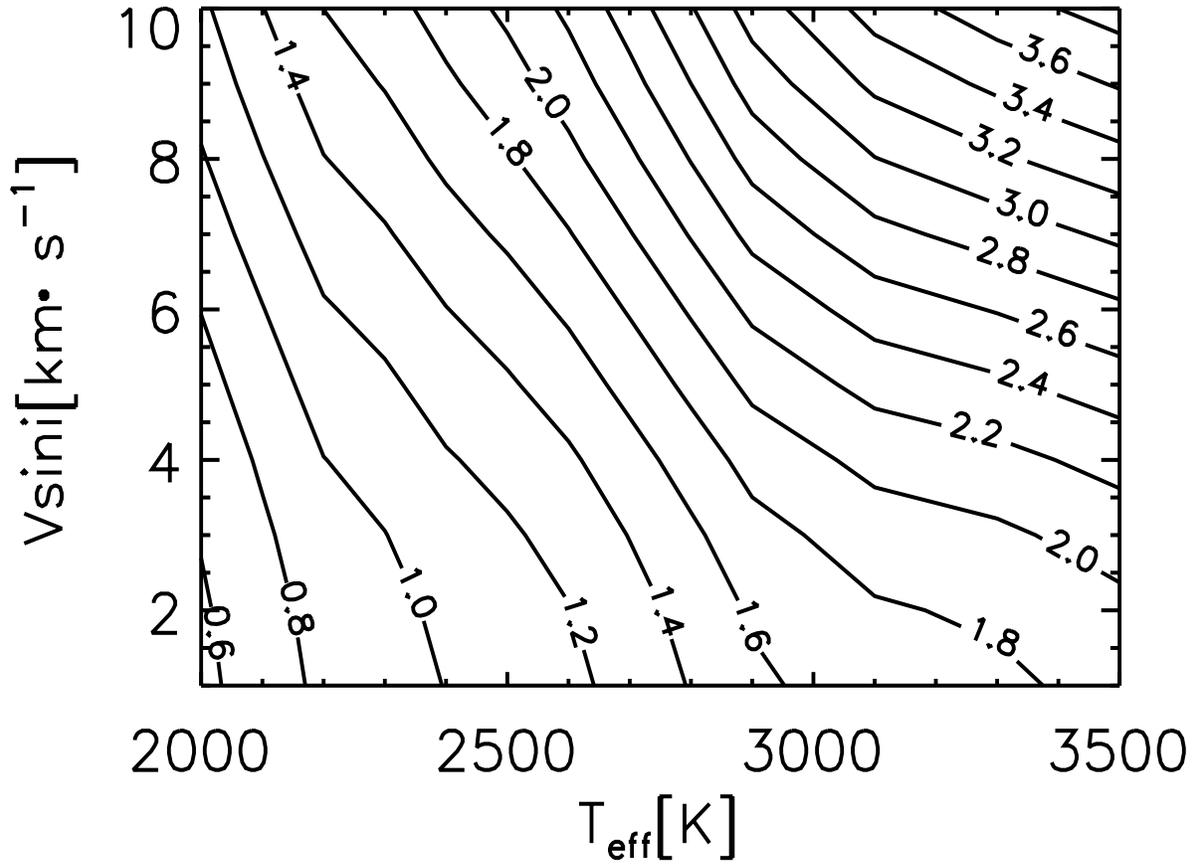} \caption{Predicted photon-limited RV uncertainty contours (in $\rm{m\cdot s}^{-1}$)
for IRET. The assumption in calculation includes: 1)
$t_{\rm{exp}}=30$ min; 2) $m_J=9$; 3) $\eta$ as shown in Fig. \ref{fig:Wav_Eta}. \label{fig:Teff_Vsini_contour}}
\end{center}
\end{figure}

\begin{figure}
\begin{center}
\includegraphics[width=16cm,height=12cm]{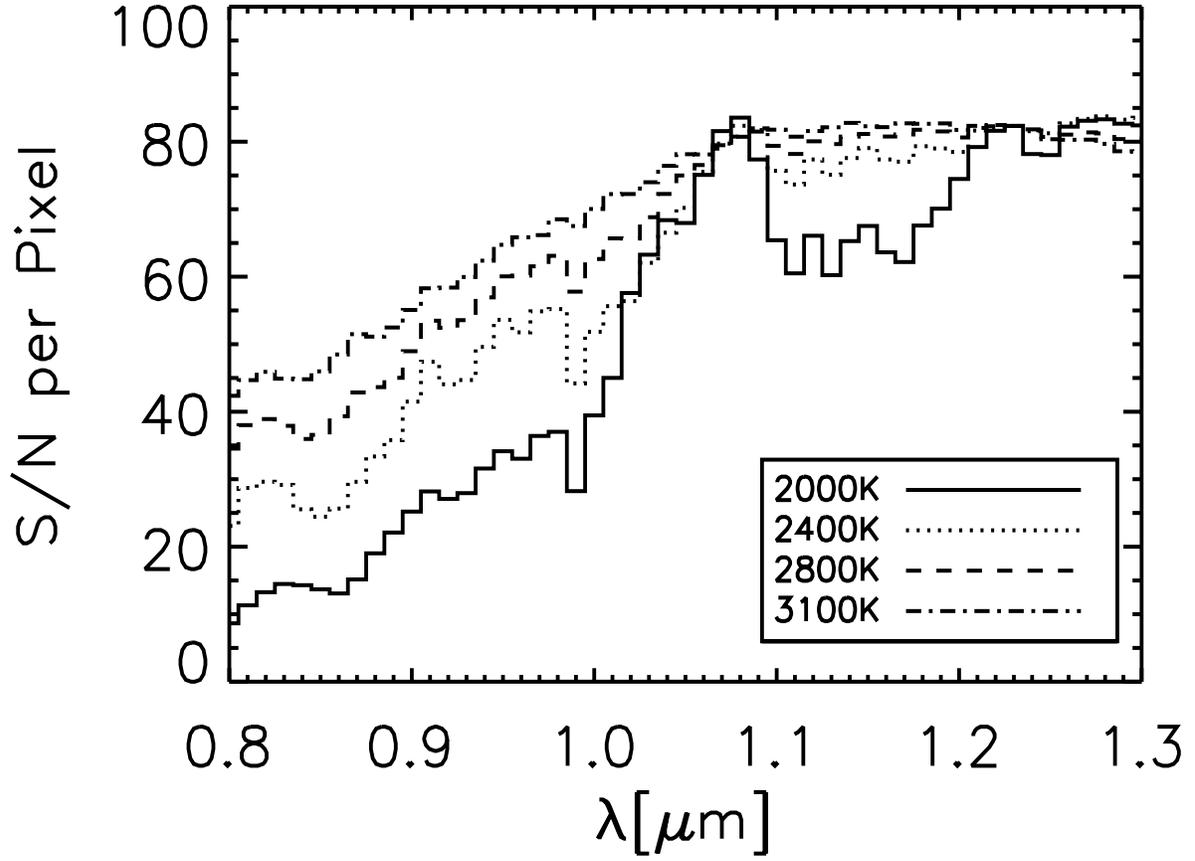} \caption{S/N per pixel as a function of wavelength. The assumption in calculation includes: 1)
$t_{\rm{exp}}=30$ min; 2) $m_J=9$; 3) $\eta$ as shown in fig
\ref{fig:Wav_Eta}. IRET is designed to have 4.2 pixels to sample one RE
and 25 pixels to sample fringes in the slit direction in each spectral channel. \label{fig:Wav_SNR}}
\end{center}
\end{figure}

We conduct a series of simulations as described in \S \ref{sec:edi}
using synthetic spectra of M dwarfs and the instrument
specifications of IRET~\citep{Zhao2010} to seek for an optimized OPD that maximizes
the $Q$ factor. Fig. \ref{fig:Q_temp} shows results from simulations
of M dwarfs at different $T_{\rm{eff}}$. It is shown that for $T_{\rm{eff}}$
ranging from 2400K to 3100K, the $Q$ factor is maximized at an OPD of
around 18.0 mm for $V \sin{i}$ between 3 $\rm{km\cdot s}^{-1}$ and 5 $\rm{km\cdot s}^{-1}$. In
addition to the $Q$ factor, we also need photon flux information for a
particular star to calculate the photon-limited RV uncertainty (Equation
(\ref{eq:overall_Doppler_2d})), which is given by Equation
(\ref{eq:photon_flux}). We calculate $F_*$ using a synthetic stellar
spectrum and then normalize it such that the flux at 1235 nm
is 195.2 $\rm{photons}\cdot s^{-1}\cdot\AA^{-1}\cdot cm^{-2}$. Instrument throughput is as shown in
Fig. \ref{fig:Wav_Eta}, which is estimated based on our current optical design~\citep{Zhao2010}. Note that the estimation does not consider fiber coupling efficiency, fiber efficiency, seeing coupling loss and atmospheric transmission. We choose the telescope aperture size of 3500 mm as the design reference to calculate $S_{\rm{tel}}$. The telescope obscuring structure is  accounted for in the total detection efficiency, $\eta$. We assume a
$t_{\rm{exp}}$ of 30 min on an $m_J=9$ star in the RV uncertainty
contours. Fig. \ref{fig:Teff_Vsini_contour} shows the fundamental photon-limited RV
uncertainty for a set of grids of $T_{\rm{eff}}$ and $V \sin{i}$. The photon noise limited
Doppler precision can reach $\sim$1-2 $\rm{m}\cdot\rm{s}^{-1}$ for slow rotating cool stellar objects. 
There is a general trend of RV uncertainty increasing from the low $T_{\rm{eff}}$ and
low $V \sin{i}$ to the high $T_{\rm{eff}}$ and high $V\sin{i}$. Average S/N per pixel as a function of wavelength is plotted in Fig. \ref{fig:Wav_SNR}.

As of June 2011, there were only six M dwarf exoplanets
discovered in the northern
hemisphere~\citep{Butler2004,Charbonneau2009,Forveille2009,Howard2010,Haghighipour2010,Johnson2010}.
We compare the velocity semi-amplitude $K$ of these exoplanets and
RV uncertainty predicted for IRET (Table \ref{tab:Detected_M_planet}). All of them would be detectable by IRET under photon-limited conditions. Therefore,
IRET is a suitable instrument conducting follow-up RV measurements of these 
targets to confirm the detection and detect additional planet companions in these
systems. We also compare the velocity
semi-amplitude, $K_{\rm{HZ}}$, of an Earth-mass planet located
within the habitable zone (0.05 AU away from the host star) to the 
IRET photon-noise detection limit and find that the RV uncertainty
of IRET is slightly larger than $K_{\rm{HZ}}$. It may be challenging to 
detect a possible Earth-mass planet in the habitable zone of these 
planet systems.  However, IRET is capable of detecting
 super-Earth exoplanets in the HZ or
Earth-like planets in the HZ around nearby brighter M dwarfs under
photon-limited condition.

Earlier study showed that NIR RV measurements gain precision
for objects later than M4 because of relatively higher flux and increasing stellar absorption features in the NIR~\citep{Reiners2010}.
In addition, our study in earlier sections showed that
 stellar rotation intrinsically broadens stellar
absorption features and results in less Doppler information, leading to 
larger RV measurement uncertainty with IRET.
Therefore, slow rotating mid-late type M dwarfs would be major science
targets for a planet survey with IRET. 

We conducted a literature survey to estimate how many
potential targets are suitable for the M dwarf survey in the northern sky 
using IRET.  We select our potential survey targets from the LSPM catalog,
which  was
 compiled by Lepine (2005) through a
photometric survey of the entire northern sky.
This catalog is nearly complete 
for stars with proper motion higher than 0.15$^{{\prime\prime}}\rm{yr}^{-1}$. 
We apply a color cut, V-J$>$3,
to select M type stars from the LSPM catalog.
Table \ref{tab:M_dwarf_targets} shows the number of M dwarf targets
available brighter than certain J band magnitude. However, not all of the 
bright M dwarfs are suitable targets. In order to reach high precision (1-3 $\rm{m}\cdot\rm{s}^{-1}$),
we only choose M dwarfs with $V \sin{i}$ less than 10 $\rm{km\cdot s}^{-1}$ 
as major targets. However, previous study shows that there appears  to be a trend that the median of $V \sin{i}$ increases as $T_{\rm{eff}}$
decreases~\citep{Jenkins2009}. According to \citet{Jenkins2009}, mid-late
type M dwarf $V \sin{i}$ distribution can be fitted by a power law
with an index of -1.12, which indicates that 41\% of mid-late M
dwarfs have a $V \sin{i}$ less than 2 $\rm{km\cdot s}^{-1}$ and 74\% of them have a
$V \sin{i}$ less than 10 $\rm{km\cdot s}^{-1}$. 
They also pointed out that 43\% of M
dwarfs in their sample exhibit an $\rm{H}\alpha$ emission that indicates
young and active objects not suitable for RV survey. Statistically,
there are about 18\% of nearby M dwarf targets that 
are suitable for a precision RV survey (inactive and with $V \sin{i}<2\ \rm{km\cdot s}^{-1}$). Based on IRET photon-limited performance, 
it is likely that IRET can reach less than 3 $\rm{m}\cdot\rm{s}^{-1}$ photon-limited Doppler
precision for over 200 nearby late M dwarfs to detect and characterize 
Earth-mass and Super-Earth mass planets.

\section{Telluric Line Effect}
\label{sec:Telluric}

\subsection{Quantitative Telluric Line Effect }

\begin{figure}
\begin{center}
\includegraphics[width=16cm,height=12cm]{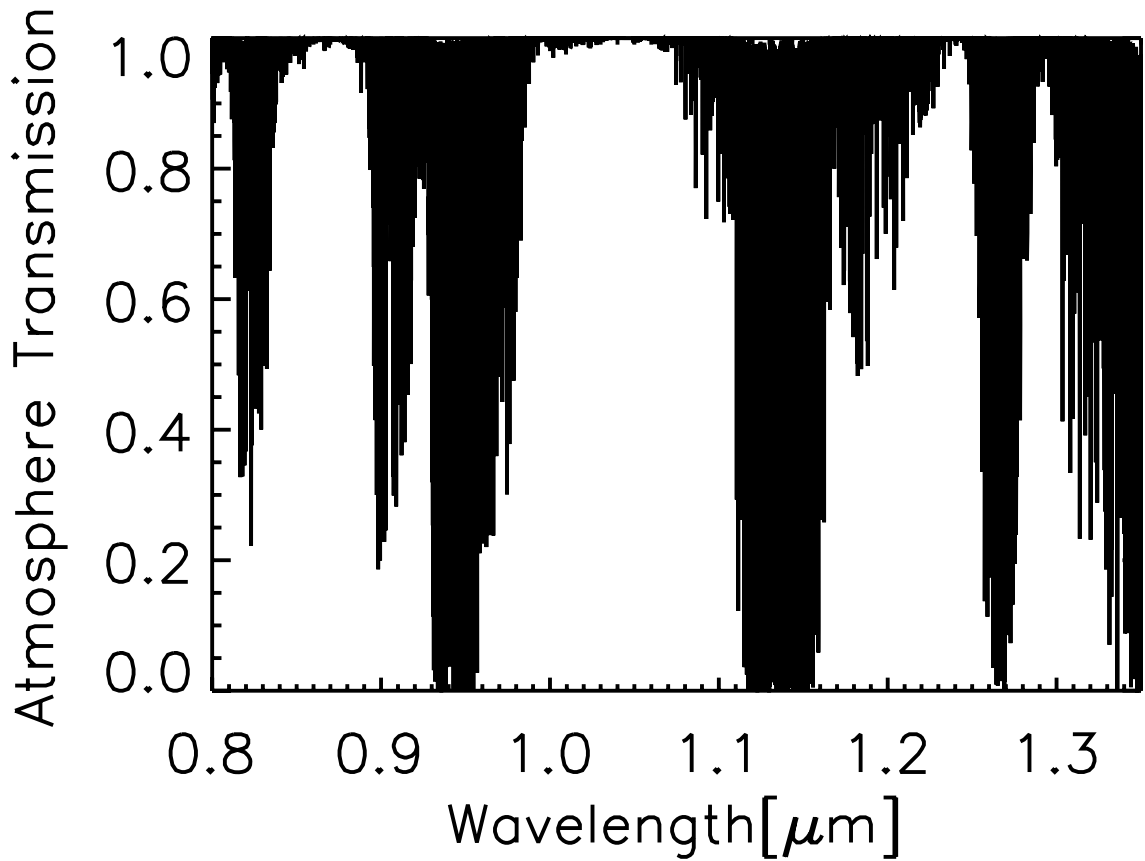}
\caption{Atmosphere transmission (AT) as a function of wavelength
(Generated by ATRAN~\citep{Lord1992}). \label{fig:ir_tr}}
\end{center}
\end{figure}

Ground-based NIR observation is prone to contamination of telluric
lines. Precision RV measurements in the NIR is an extreme case that requires significant disentanglement of stellar absorption
lines and telluric lines in order to reach photon-limited precision.
 Even in the Y band (960 to 1080 nm)
and J band (1120 to 1320 nm), where telluric line contamination
is less severe, there are still many shallow telluric lines
(Fig. \ref{fig:ir_tr},~\cite{Lord1992})
\footnotemark[1]\footnotetext[1]{http://atran.sofia.usra.edu/cgi-bin/atran/atran.cgi}.
Equation (\ref{eq:B_no_atm}) should be rewritten as follows if
telluric absorption lines are considered:
\begin{eqnarray}
B(\nu,y) & = & \bigg[\frac{S_0(\nu)}{h\nu}\times AT(\nu) \times
IT(\nu,y)\bigg]\otimes LSF(\nu,R) \nonumber \\
   & = & \bigg[\frac{S_0(\nu)}{h\nu}\times (1-AA(\nu)) \times
IT(\nu,y)\bigg]\otimes LSF(\nu,R) \nonumber \\
   & = & \bigg[\frac{S_0(\nu)}{h\nu}\times IT(\nu,y)\bigg]\otimes LSF(\nu,R)+\bigg[-\frac{S_0(\nu)}{h\nu}\times AA(\nu) \times
IT(\nu,y)\bigg]\otimes LSF(\nu,R) \nonumber \\
   & = & B_S(\nu,y)+B_N(\nu,y),
\label{eq:B_atm}
\end{eqnarray}
where $AT$ is the atmospheric transmission function and $AA$ is the
atmospheric absorption function. In Equation (\ref{eq:B_atm}),
photon flux distribution on the detector, $B$, is comprised of a
signal component $B_S$ and a noise component $B_N$. Ideally, we
require that the detector flux change $\delta B$ is entirely due to the
stellar RV change $\delta v_S$. However, $\delta B$ is also partly
induced by telluric line shift $\delta v_N$ resulting from
atmospheric behaviors. Therefore, both $\delta v_S$ and $\delta v_N$
contribute to $\delta B$. We have two sets of RV measurements,
$\delta v_S+\sigma(0,\delta v_{rms,S})$ for RV of $B_S$ and $\delta
v_N+\sigma(0,\delta v_{rms,N})$ for RV of $B_N$, where
$\sigma(0,\delta)$ represents random numbers following a gaussian
distribution with a mean of 0 and a standard deviation of $\delta$.
We weight RV measurements with the inverse square of photon-limited
RV uncertainties. The final RV measurement is expressed by the
following equation:
    \begin{equation}
    \label{eq:simple_example_tulleric}
\delta v=\frac{(\delta v_S+\sigma(0,\delta
v_{rms,S}))\cdot{\delta v_{rms,S}^{-2}}+(\delta
v_N+\sigma(0,\delta v_{rms,N}))\cdot{\delta
v_{rms,N}^{-2}}}{{\delta v_{rms,S}^{-2}}+{\delta
v_{rms,N}^{-2}}}.
\end{equation}
In practical Doppler measurements, $\delta v_S$ consists two components, stellar RV and Earth's barycentric RV. We assume a constant stellar RV (i.e., no differential stellar RV). The Earth's barycentric motion may be predetermined and the effect is removed from the observed 2-D spectrum by shifting $B_S$ according to the barycentric velocity. Therefore $\delta v_S$=0 for the $B_S$ component. We further assume that $B_N$ has an RV fluctuation of $\sigma(0,\delta v_{N,ATM})$ because of the Earth's  turbulent atmosphere. The measured RV
uncertainty $\delta v$ is equal to:
    \begin{equation}
    \label{eq:simple_example_tulleric_2}
\delta v_{rms}=\frac{(\delta v_{rms,S})\cdot{\delta
v_{rms,S}^{-2}}+{(\delta v_{N,ATM}^2+\delta
v_{rms,N}^2)^{1/2}}\cdot{\delta v_{rms,N}^{-2}}}{{\delta
v_{rms,S}^{-2}}+{\delta v_{rms,N}^{-2}}},
\end{equation}
Three examples are considered here to represent different level of 
telluric contamination to the precision RV measurements: 1, if $\delta v_{rms,S}\gg\delta v_{rms,N}$, then $\delta v_{rms}={(\delta v_{N,ATM}^2+\delta
v_{rms,N}^2)^{1/2}}$, and the RV measurement is dominated by atmospheric behaviors, this approximation applies in a wavelength region with dense telluric line distribution;
2, if $\delta v_{rms,S}\ll\delta v_{rms,N}$ (i.e., in very transparent atmosphere windows where few telluric lines exists), then $\delta
v_{rms}=\delta v_{rms,S}$, and the RV uncertainty is limited by stellar
photon noise; 3, for an intermediate situation, if $\delta v_{rms,S}$ and $\delta
v_{rms,N}$ are identical, then we apply the same weight on both RV
measurements, $\delta v_{rms}={[\delta v_{rms,S}+{(\delta
v_{N,ATM}^2+\delta v_{rms,N}^2)^{1/2}}]/2}$.

In the practical NIR spectroscopic 
observations, telluric contamination must be minimized 
in order to reach high Doppler precision. We investigate the effectiveness of 
two major ways to remove telluric contamination from the stellar spectra with IRET:
 telluric line masking and removing, and telluric line 
modeling and removing. The study results are summarized below. 

\subsection{Telluric Line Masking}
Telluric absorption lines are not homogeneously distributed in the NIR 
spectra, instead, a large number of telluric lines are
 from relatively concentrated spectral regions. 
The simplest way  is to mask and remove those severely
contaminated regions while leaving the less contaminated stellar lines for 
RV measurements. This would reduce RV measurement uncertainty caused by
telluric line contamination, leading to improved RV precision.
However, if  too much of the wavelength coverage
region is masked and removed, then RV uncertainty due to photon noise would increase. Therefore, a balance between the uncertainty brought
by telluric lines contamination and by photon noise must be found and 
an optimal RV performance can be achieved.

We use an M9 dwarf ($T_{\rm{eff}}$=2400K, $V \sin{i}$=5$\rm{km\cdot s}^{-1}$ and $m_J$=9) as an example to illustrate how this masking technique 
affects the RV measurement precision with IRET. We assume a 30 min exposure time, a  wavelength coverage from 800 to 1350 nm 
and an instrument throughput as shown in Fig. \ref{fig:Wav_Eta}. 
We  calculate photon-limited RV
uncertainty $\delta v_{S,rms}$ and $\delta v_{N,rms}$ according to
Equation (\ref{eq:overall_Doppler_2d}).
We calculate $Q$ factors, $Q_S$ and $Q_N$, based on the two components, $B_S$ and
$B_N$, in Equation (\ref{eq:B_atm}).  The photon flux of $B_S$ and $B_N$ are 
calculated from the inputs of the stellar type,
magnitude, exposure time, instrument specifications and telluric
absorption properties.

In practice, RV uncertainty of $B_N$ is not
dominated by photon-noise, instead, it is dominated by atmospheric
behaviors such as wind, molecular column density change, etc. ~\citet{Figueira2010} used HARPS archive data and found
that $O_2$ lines are stable to a 10 $\rm{m\cdot s}^{-1}$ level over 6 years. However,
the stability of telluric lines becomes worse if we take into
consideration other gas molecules such as $H_2O$ and $CO_2$. We
assume different levels of RV fluctuation due to atmospheric
behaviors $\delta v_{N,ATM}$ (5 $\rm{m\cdot s}^{-1}$, 10 $\rm{m\cdot s}^{-1}$ and 20 $\rm{m\cdot s}^{-1}$). The
uncertainty induced by atmospheric telluric lines is transferred to
$\delta v_{rms}$ via Equation (\ref{eq:simple_example_tulleric_2})
without the effort of disentangling stellar and telluric lines as if
the uncertainty is due to photon-noise of $B_S$. Telluric lines with
the largest slope are masked at the highest priority because
they cause the most RV uncertainty in the measurement.

\begin{figure}
\begin{center}
\includegraphics[width=16cm,height=12cm]{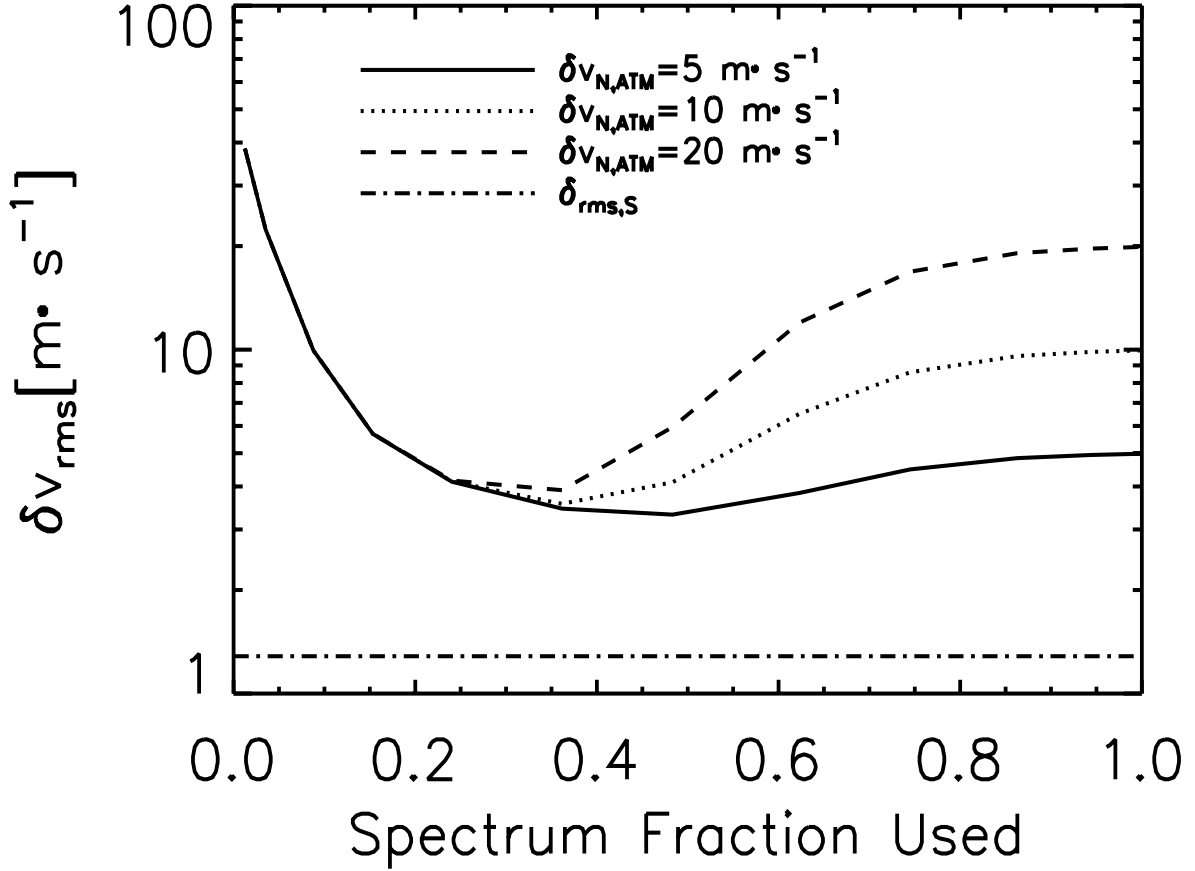} \caption{Predicted RV uncertainty $\delta v_{rms}$ as a function
of the fraction of stellar spectrum used in RV measurements when applying the
telluric line masking technique. We assume $T_{\rm{eff}}$=2400 K and $V \sin{i}$
of 5 $\rm{km\cdot s}^{-1}$. Different levels of atmosphere RV fluctuation are indicated by different line styles. $\delta v_{rms,S}$ (dash-dotted line), the fundamental photon-limited RV uncertainty determined by an intrinsic stellar spectrum for a complete wavelength coverage from 800 to 1350 nm, is plotted for comparison. 
\label{fig:RV_Un_Mask}}
\end{center}
\end{figure}

Fig. \ref{fig:RV_Un_Mask} shows the results of telluric line masking for
an M9 dwarf under
different levels of telluric line masking. After masking a certain
portion of the most severely contaminated stellar spectrum, optimal
RV uncertainty $v_{rms}$ is 3.3 $\rm{m\cdot s}^{-1}$, 3.6 $\rm{m\cdot s}^{-1}$ and 3.9 $\rm{m\cdot s}^{-1}$
respectively for $\delta v_{N,ATM}$ of 5 $\rm{m\cdot s}^{-1}$, 10 $\rm{m\cdot s}^{-1}$ and 20 $\rm{m\cdot s}^{-1}$. They are 2.6, 2.8 and 3.0
times worse compared to the fundamental photon-limited RV uncertainty of the stellar
spectrum $\delta v_{rms,S}$, i.e., 1.3 $\rm{m\cdot s}^{-1}$. If we do not apply telluric
line masking at all, RV uncertainty is dominated by atmospheric
behaviors, $\delta v_{N,ATM}$. It is expected that $\delta v_{rms}$
increases as we mask more of the stellar spectrum because less
photon flux is available to calculate RV. Depending on values of
$\delta v_{N,ATM}$, the optimal spectrum fraction used in RV
measurement lies in between 30\% and 50\%. Higher value of
$\delta v_{N,ATM}$ requires masking and removing larger fraction of stellar spectrum. It is noted that we would have expected $\delta v_{rms}$
to be only $\sim$1.4 times worse than $\delta v_{rms,S}$ after $\sim$50\%
of the spectrum is masked for pure
concern of photon flux. However, $\delta v_{rms}$ is $\sim$3 times
worse than $\delta v_{rms,S}$ because
not only  $N_{e^-}$ but also $Q$ is reduced after telluric line
masking.

\subsection{Telluric Lines Modeling}
\label{sec:TelluricRemoval}
In principle, the telluric lines can be modeled and removed using a telluric
line spectrum template. This kind of template can be obtained by observing 
fast rotating early type stars since
 the spectrum of an early type star (early A type
in particular) is nearly featureless and a good approximation of a blackbody spectrum in
NIR except for broad hydrogen absorption features\citep{Vacca2003}.
However, the number of available telluric standard stars limits the
application of this method. In addition, the spectrum of a telluric standard
star is not entirely featureless, which results in some residuals that undermines
precision RV measurements. 

As a first order estimation, we consider the 
simplest case in which we assume that the
spectrum of a telluric standard star is a perfect blackbody spectrum
with no spectral features which may otherwise be perfectly modeled and
removed. Since the spectrum of a science target and the spectrum of
a telluric line standard star are not taken simultaneously, the
telluric lines' centroids and depths may change over time. We assume
that the centroids of telluric lines remain unshifted. As a result, the only
uncertainty is the depth of a telluric line. Equation
(\ref{eq:overall_Doppler_2d}) and (\ref{eq:B_atm}) offer insight in
understanding the process of telluric line modeling. In the telluric
line modeling, we model the telluric lines by observing a standard
star or by forward modeling using a  list of telluric lines. We remove
telluric lines from the observed spectrum of a science target. The depths of
the telluric lines are consequently reduced by a certain fraction
depending upon the level of removal, which effectively
reduces $AA$ (Atmosphere Absorption). A reduced $AA$ means less
photon contribution in $B_N$ and thus less uncertainty induced by atmospheric behaviors. 

\begin{figure}
\begin{center}
\includegraphics[width=16cm,height=12cm]{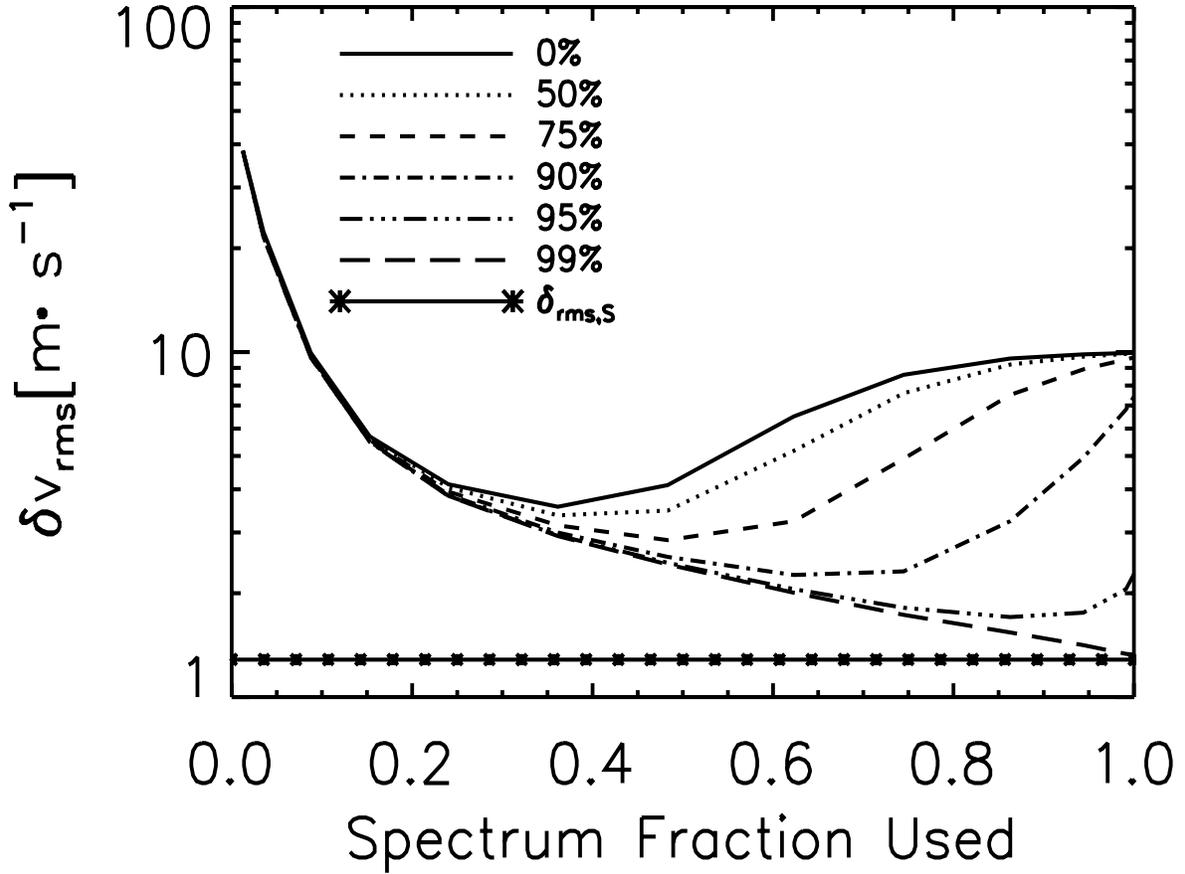} \caption{Predicted RV uncertainty as a function of fraction of
stellar spectrum used in RV measurements when applying the telluric
line  modeling  and removing technique. We assume $T_{\rm{eff}}$=2400 K and $V
\sin{i}$ of 5 $\rm{km\cdot s}^{-1}$. Asterisks connected by solid line represent the
fundamental photon-limited RV uncertainty. Different line styles
represent different level of telluric lines removal
(i.e, removed telluric line strength). $\delta v_{N,ATM}$=10 $\rm{m}\cdot\rm{s}^{-1}$ is assumed
in the calculations.\label{fig:RV_Un_Mask_Removal}}
\end{center}
\end{figure}

\begin{figure}
\begin{center}
\includegraphics[width=16cm,height=12cm]{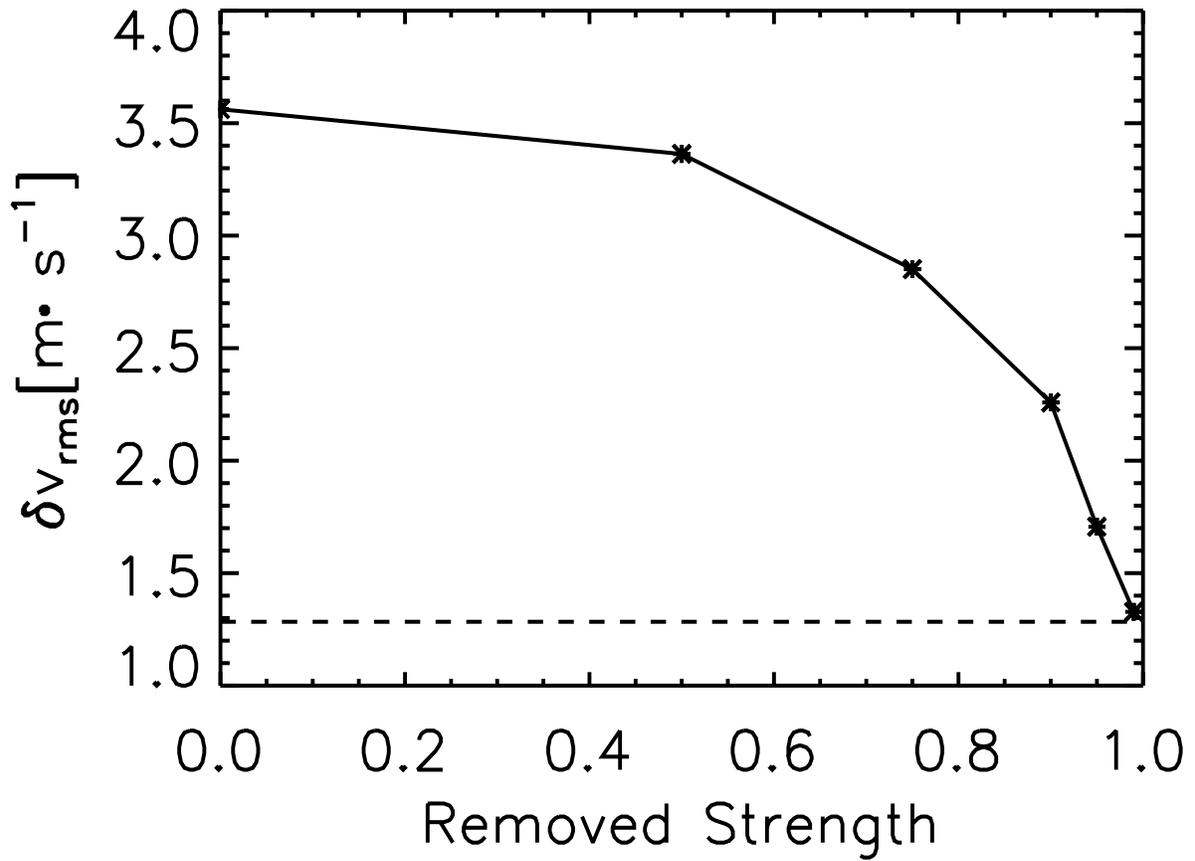} \caption{Predicted optimal RV uncertainty as a function of removed
telluric line strength. Dashed line represents the fundamental
photon-limited RV uncertainty, $v_{rms,S}$. \label{fig:RV_Un_Strength_Removed}}
\end{center}
\end{figure}

Fig.
\ref{fig:RV_Un_Mask_Removal} shows how RV uncertainty $\delta
v_{rms}$ is correlated with different levels of telluric line modeling
 and removing. We assume $\delta v_{N,ATM}$ to be 10 $\rm{m\cdot s}^{-1}$. Fig.
\ref{fig:RV_Un_Strength_Removed} shows the optimal RV uncertainty at different
removal levels. If no removal is involved, the optimal
 $\delta v_{rms}$ is 3 times worse than
$\delta v_{rms,S}$. At low levels of
telluric line removal, i.e., a small fraction of the strength of
telluric lines is removed, it requires to  mask
out about 30\% to 50\% of the stellar spectrum to reach an optimal
RV uncertainty. As the removal becomes more effective,
the influence caused by telluric lines decreases, and only a small portion of the spectrum requires masking. Therefore, the optimal RV
uncertainty, $\delta v_{rms}$, reaches a smaller value. At a very high level of removal, in which 99\% of the strength of telluric lines
can be modeled and removed, i.e., the residual of telluric line
modeling is only 1\%, telluric line masking is not necessary. The entire
wavelength coverage region can be used in the RV calculation. $\delta
v_{rms}$ approaches the fundamental photon-limited RV uncertainty,
$\delta v_{rms,S}$. In previous study, the residual of telluric lines 
has reached an rms of 0.7\% using a 3-component
(absorption gas, telluric lines and star) spectrum to model the observed
stellar spectrum~\citep{Bean2010}. Therefore, it is quite possible 
to reach photon noise limited RV precision with IRET using this telluric 
line modeling and removing method.

\section{Summary and Discussion}
\label{sec:Conclusion}

\subsection{$Q$ Factors for DFDI, DE and FTS}

We develop a new method of calculating photon-limited Doppler
sensitivity of an instrument adopting the DFDI method. We conduct a series of
simulations based on high resolution synthetic stellar spectra generated
by PHOENIX code\citep{Hauschildt1999,Allard2001}. In simulations, we
investigate the correlations of $Q$ and other parameters such as
OPD of the interferometer, spectral resolution $R$ and stellar
projected rotational velocity $V \sin{i}$. We find that optimal OPD
increases with increasing $R$ and decreasing $V \sin{i}$. Empirically, the optimal OPD is chosen such that the density of the interference combs matches with the line density of the stellar spectrum. Based on the simulation results, the optimal OPD is determined as the one that maximizes the $Q$ factor. In fact, optimal OPDs found from empirical way and from numerical simulation  are consistent with each other. For example, for $V\sin i$=0 $\rm{km}\cdot\rm{s}^{-1}$ and $R$=50,000, simulation gives an optimal OPD of 30 mm. The interference comb density of an interferometer with OPD of 30 mm is $\sim$0.3 \AA\ at 1000 nm, which indeed matches the width of a typical absorption line after spectral blurring with $R$ of 50,000. An independent method to calculate photon-noise limited Doppler measurement uncertainty in the optical is being developed, and the results will be reported in a separate paper~\citep{Jiang2011}. We have compared results from both methods and confirmed that both independent methods produce essentially the same results for both optical and NIR Doppler measurements. 

We investigate how the $Q$ factor is affected if OPD is deviated from the
optimal value and find that a deviated OPD (5mm) does not result
in a significant $Q$ factor degradation, which is mitigated as $R$
increases. We find that the $Q$ factor increases with $R$ for both
DFDI and DE, and eventually converge at very high $R$ ($R\geq$100,000). The convergence of DFDI and DE methods is a natural consequence because the measurement method does not make a difference after the spectral resolution becomes extremely high. In addition, $Q$ factors at a given $R$ increase as $T_{\rm{eff}}$ drops from 3100K to 2400K, which is due to stronger molecular absorption
features in NIR (see Fig. \ref{fig:Wav_Flux}). The $Q$ factor
decreases as $V \sin{i}$ increases because stellar rotation broadens the absorption lines, leading to less sensitive measurement.

We compare $Q$ factors for both DFDI and DE at a given $R$. For slow
rotators ($0\ \rm{km\cdot s}^{-1}\le V \sin{i}\le2\ \rm{km\cdot s}^{-1}$), DFDI is more
advantageous over DE at low and medium $R$ (5,000 to 20,000) for the same
wavelength coverage $\Delta\lambda$. The improvement of DFDI
compared to DE is $\sim$3.1 ($R$=5,000), $\sim$2.4 ($R$=10,000) and
$\sim$1.7 ($R$=20,000), respectively. In other words, optimized DFDI
with $R$ of 5,000, 10,000 and 20,000 are equivalent in Doppler
sensitivity to DE with $R$ of 16,000, 24,000 and 34,000, respectively.
The improvement of DFDI at $R$ 20,000 to 50,000 is not as noticeable as at
 low $R$ range. The difference between DFDI and DE becomes
negligible when $R$ is over 100,000. For relatively faster rotators
($5\ \rm{km\cdot s}^{-1}\le V \sin{i}\le10\ \rm{km\cdot s}^{-1}$), the improvement with DFDI is less obvious than it is for very slow rotators. DFDI has strength when the spectral lines in a stellar spectrum are not resolved by a spectrograph, which is the case for low and medium resolution spectrograph. Under such conditions, the fixed delay interferometer provides additional resolving powers for the system. After the lines are fully resolved by the spectrograph itself, the interferometer in the system becomes dispensable, which is the reason why we see the convergence of DFDI and DE at very high spectral resolution. 

Fundamental performance of a Fourier-transform spectrometer (FTS) in the application of Doppler measurements has been discussed by~\citet{Maillard1996}. There are similarities between the FTS and the DFDI method, for example: 1, both methods use the interferometer as a fine spectral resolving element; 2, RV is measured by monitoring the temporal phase change at a fixed OPD of the interferometer. In DFDI method, OPD is scanned in each frequency channel because of two relatively tilted mirrors, and the resolution of the post-disperser in DFDI is chosen to ensure a reasonable fringe visibility. Therefore, the DFDI method is a extended version of the FTS method with a low-medium resolution post-disperser. However, one major difference between these two methods is that the interferometer itself is used as a spectrometer by OPD scanning in the FTS method while an additional spectrograph is employed in the DFDI method. The advantage of introducing an additional spectrograph into the system is that the visibility (or fringe contrast) is no longer limited by the bandpass as in the FTS case, which is the reason that the DFDI method can be applied in broad-band Doppler measurements. ~\citet{Mosser2003} discussed the possibility of an FTS working in broad band by introducing a low resolution post-disperser and concluded that the FTS method is inferior (by a factor between 1 and 2) to DE method even after employing an  post-disperser. This conclusion should be accepted with cautions because they compared an FTS with a post-disperser ($R$=1200) with a DE instrument with a much higher spectral resolution ($R$=84,000), which is not necessarily a fair comparison. 


We define new merit functions (Equations
(\ref{eq:Qprime1}) and (\ref{eq:Qprime2})) to objectively
 evaluate Doppler performance for both DFDI and DE methods. For $Q^\prime$, the merit function for single object observation, we find that
$Q^\prime_{\rm{DFDI}}$ is consistently higher than $Q^\prime_{\rm{DE}}$
regardless of the $R$ of the DE instrument under the constraint of total number of pixels, i.e., both the DFDI and DE
instrument adopt the same NIR detector. The
DE instrument requires using a larger detector in order to reach the
same wavelength coverage as the DFDI instrument. Note that the above conclusion is based on the assumption that the number of pixels per spectral order are the same for DFDI and DE. In practice, a DFDI instrument uses $\sim$20 pixels to sample spatial direction, i.e., the direction transverse to dispersion direction, while $\sim$5 pixels are usually used to sample spatial direction in a DE instrument. However, the $\sim$20 pixels sampling is not a requirement for DFDI but rather for the convenience of data reduction. Normally, $\sim$7 pixels sample one spatial period of a stellar fringe, which in principle are adequate based on a phase-stepping algorithm provided by ~\citet{Erskine2003}.

If the same detector is used, the spare part of the detector in DFDI can be used for multi-object observations. Consequently, in addition to single-object instrument, we also investigate $Q^{\prime\prime}$, a merit function for multi-object RV measurement for both DFDI and DE. Different conclusions are reached depending on different value of $\alpha$, an index of the importance of multi-object observation. From a pure photon gain point of view, DFDI and DE instruments have similar $Q^{\prime\prime}$ values with $Q^{\prime\prime}_{\rm{DFDI}}$ slightly better than $Q^{\prime\prime}_{\rm{DE}}$ (a factor of $\sim$1.1). From a survey efficiency point of view, a DFDI multi-object instrument is 9 times faster than its counterpart using DE for slow rotating stars ($0\ \rm{km\cdot s}^{-1}\le V \sin{i}\le2\ \rm{km\cdot s}^{-1}$) and $\sim$4 times faster for fast rotators ($V \sin{i}\ge10\ \rm{km\cdot s}^{-1}$).

\subsection{Application of DFDI}

InfraRed Exoplanet Tracker (IRET) is used as an example to demonstrate 
RV performance with the DFDI method and illustrate how the RV performance
is related to other parameters. It
is shown that for slow rotators with $T_{\rm{eff}}$ ranging from 2400K to
3100K, the $Q$ factor is maximized at an OPD of around 18.0 mm. According to predicted photon-limited RV precision, IRET is capable of detecting 
Earth-like planets in habitable zone around bright M dwarfs if they exist. In our simulations, we consider the photon loss due to optics, CCD detector and the telescope transmission loss of 20\%. For the real observation, the total detection 
efficiency needs to include the atmospheric transmission loss,  fiber coupling and seeing losses. 
It is likely that the overall detection efficiency is a factor of two lower, which 
requires 2 times longer exposure to reach the same Doppler precision. In addition, the real instrument 
may have short and long term systematics which affect RV measurement precision. 
 Therefore, the performance of IRET predicted in the paper should be considered 
as an estimation in an optimistic case. 

There may be other practical concerns about the instrument using the DFDI method, most of them are due to the relative low spectral resolution compared to current DE instruments. First of all, an absolute wavelength calibration for a DFDI instrument is not as precise as a DE instrument with a higher spectral resolution. For example, at a spectral resolution of 22,000 for IRET, a line profile with a FWHM of $\sim$0.45 $\AA$ in $Y$ band is expected. Following the method described in~\citet{Butler1996}, it corresponds to 136.4 $\rm{m\cdot s}^{-1}$ RV uncertainty at a S/N of 100 if only one spectral line is used.~\citet{Ramsey2010} proposed to use a U-Ne emission lamp as a wavelength calibration source and it has approximately $\sim$500 lines in $Y$ band according to their measurement. Therefore, after all the lines in $Y$ band are considered, $\sim$6 $\rm{m\cdot s}^{-1}$ RV uncertainty is introduced in the process of absolute wavelength calibration. In comparison, a DE instrument at $R$ of 110,000 causes $\sim$1.2 $\rm{m\cdot s}^{-1}$ RV uncertainty in an absolute wavelength calibration. However, an absolute wavelength solution is only required for the DE method in order to measure RV drift due to instrument instability, which is measured in a different method in a DFDI instrument. It is similar to a stellar RV measurement, the difference is that the object is switched from a star to an wavelength calibration source. Vertical fringe movement of absorption or emission lines of an RV calibration source is measured instead of centroid movement measurement in a DE instrument. In this case, a DFDI instrument (e.g., IRET, $R$=22,000) is equivalent to a DE instrument with $R$ of 37,000 in terms of Doppler measurement precision (see \S \ref{sec:DFDI_Res}). Therefore, instrument RV drift calibration process introduces an RV uncertainty of $\sim$3.5 $\rm{m\cdot s}^{-1}$ for IRET in the example of a U-Ne lamp calibration source. In addition, the RV uncertainty can be further reduced by increasing S/N and number of measurement. Secondly, at a low spectral resolution, it is challenging to perform spectral line profile analysis and thus it requires high-resolution follow-up in order to confirm or exclude a possible detection. Last but not least, in a binary case in which the observed spectrum is blended, two approaches can be used for identification: 1, from measured RV, if the flux ratio is small, similar to the planet companion case, then the lower mass companion can be identified in the measured RV curve even though small flux contamination exists; if the flux ratio is about unity, indicating strong flux contamination, a large RV scattering is expected because this case is not considered and modeled in the current data reduction pipeline; 2, from measured spectrum, even the observed spectrum is a 2-D fringing spectrum in DFDI, we can still de-fringe the spectrum into a 1-D traditional spectrum, on which special treatment can be performed to quantify the blending such as TODCOR~\citep{Zucker2003}.

\subsection{Telluric Contamination}

We develop a quantitative method of estimating telluric line
influence on RV uncertainty. We confine the discussion within the context of DFDI method, but it can be readily applied to DE method. We assume different levels of RV
fluctuation due to atmospheric behaviors $\delta v_{N,ATM}$ (5 $\rm{m\cdot s}^{-1}$,
10 $\rm{m\cdot s}^{-1}$ and 20 $\rm{m\cdot s}^{-1}$). After masking a certain portion of the severely contaminated stellar spectrum, optimal RV uncertainty $v_{rms}$ is
$\sim$3 times worse than the photon-limited RV uncertainty of
the stellar spectrum $\delta v_{rms,S}$. At low levels of telluric lines modeling and removing, it requires to mask out about 30\% to 50\% of a stellar
spectrum to reach an optimal RV uncertainty. As telluric line
removal becomes more effective, optimal RV uncertainty, $\delta
v_{rms}$, reaches a  smaller value. At a very high level of removal,
in which 99\% of the strength of telluric lines is removed, i.e., the residual of telluric line modeling is only 1\%,
telluric line masking becomes unnecessary. The entire wavelength
coverage region can be used in RV measurements, and $\delta v_{rms}$
approaches the fundamental photon-limited RV uncertainty, $\delta
v_{rms,S}$.

We  acknowledge the support from NSF with grant NSF AST-0705139, NASA with
grant NNX07AP14G (Origins), UCF-UF SRI program, DoD ARO Cooperative Agreement W911NF-09-2-0017, 
Dharma Endowment Foundation and the University of Florida. We thank Mr. Scott Fleming for proofreading this paper draft.

\bibliographystyle{apj}
\bibliography{mybib_JW_JG}{}



\clearpage

\begin{deluxetable}{ccccccccccccc}


\tabletypesize{\tiny}

\tablewidth{465pt}



\tablecaption{OPD choice as a function of $R$ and $V \sin{i}$ at
different $T_{\rm{eff}}$}




\tablehead{\colhead{$R$} & \colhead{} & \colhead{} & \colhead{} & \colhead{} & \colhead{} & \colhead{$V \sin{i}$} & \colhead{[$\rm{km\cdot s}^{-1}$]} & \colhead{} & \colhead{} & \colhead{} & \colhead{} & \colhead{} \\

\colhead{} & \colhead{0} & \colhead{1} & \colhead{2} & \colhead{3} &
\colhead{4} & \colhead{5} & \colhead{6} & \colhead{7} & \colhead{8}
& \colhead{9} & \colhead{10} & \colhead{} }

\startdata \label{tab:OPD_choise}
5,000 & 19,27,29 & 19,19,28 & 19,19,19 & 15,15,17 & 15,15,15 & 13,13,13 & 11,12,12 & 10,11,11 & 10,10,10 & 10,10,10 & 10,10,10 &  \\

10,000 & 21,23,27 & 19,21,26 & 19,19,20 & 17,17,17 & 15,15,15 & 13,13,13 & 11,12,12 & 11,11,11 & 10,10,10 & 10,10,10 & 10,10,10 &  \\
15,000 & 21,23,26 & 21,22,23 & 19,20,20 & 17,17,17 & 15,15,15 & 14,14,14 & 13,13,13 & 12,11,11 & 11,11,11 & 11,10,10 & 10,10,10 &  \\

20,000 & 22,23,26 & 21,23,24 & 20,21,21 & 19,19,19 & 17,17,17 & 15,15,15 & 14,14,14 & 13,13,13 & 12,12,12 & 11,11,11 & 11,11,11 &  \\
25,000 & 23,24,26 & 23,24,25 & 21,22,22 & 19,20,20 & 19,18,18 & 17,17,16 & 16,15,15 & 14,14,14 & 14,13,13 & 14,13,12 & 12,12,12 &  \\

30,000 & 24,26,26 & 24,25,26 & 23,23,24 & 21,21,21 & 19,19,19 & 18,17,17 & 17,16,16 & 16,15,15 & 14,14,14 & 14,14,14 & 14,14,14 &  \\
35,000 & 26,26,28 & 26,26,26 & 24,24,25 & 22,22,23 & 21,21,20 & 19,19,19 & 19,18,17 & 17,17,17 & 16,16,16 & 16,16,16 & 16,16,15 &  \\

40,000 & 27,28,28 & 27,27,28 & 26,26,26 & 24,24,24 & 22,22,22 & 21,20,20 & 19,19,19 & 19,18,18 & 19,18,18 & 19,18,18 & 19,18,16 &  \\
45,000 & 29,29,30 & 28,28,29 & 27,27,28 & 25,25,26 & 24,23,23 & 22,22,21 & 22,20,20 & 22,20,20 & 22,20,18 & 22,18,18 & 22,18,18 &  \\

50,000 & 30,30,31 & 29,30,30 & 28,28,28 & 27,26,26 & 24,24,24 & 24,22,23 & 22,22,22 & 22,22,22 & 22,22,22 & 22,22,22 & 24,22,22 &  \\
55,000 & 32,32,32 & 31,31,32 & 30,30,30 & 27,28,28 & 27,26,26 & 24,24,24 & 24,24,24 & 24,24,22 & 24,24,22 & 24,24,22 & 24,24,24 &  \\

60,000 & 32,32,34 & 32,32,33 & 32,31,32 & 29,28,28 & 27,27,26 & 27,26,26 & 24,24,24 & 24,24,24 & 24,24,24 & 24,24,24 & 24,24,24 &  \\
65,000 & 34,34,35 & 34,34,34 & 32,32,32 & 32,30,30 & 29,28,28 & 27,27,26 & 27,26,26 & 27,24,26 & 24,24,26 & 24,24,24 & 24,24,24 &  \\

70,000 & 35,35,36 & 35,35,36 & 34,32,34 & 32,32,32 & 32,30,30 & 32,28,28 & 27,28,28 & 27,28,26 & 27,24,26 & 24,24,26 & 24,24,26 &  \\
75,000 & 37,36,37 & 37,36,37 & 35,35,35 & 32,32,32 & 32,32,32 & 32,32,32 & 32,32,28 & 32,32,28 & 32,32,32 & 32,32,32 & 32,24,32 &  \\

80,000 & 37,37,39 & 37,37,37 & 37,36,36 & 35,34,35 & 35,32,32 & 32,32,32 & 32,32,32 & 32,32,32 & 35,32,32 & 35,32,32 & 35,32,32 &  \\
85,000 & 40,39,41 & 40,39,39 & 37,37,37 & 37,36,36 & 35,32,32 & 35,32,32 & 35,32,32 & 35,32,32 & 35,32,32 & 35,32,32 & 35,32,32 &  \\

90,000 & 40,41,41 & 40,41,41 & 40,37,39 & 37,36,36 & 37,36,36 & 37,36,36 & 35,32,36 & 35,32,36 & 37,37,36 & 37,37,36 & 37,37,36 &  \\
95,000 & 41,41,41 & 41,41,41 & 40,41,41 & 40,37,37 & 37,37,36 & 37,37,36 & 37,37,36 & 37,37,36 & 37,37,37 & 37,37,37 & 37,37,37 &  \\

100,000 & 41,41,41 & 41,41,41 & 40,41,41 & 40,40,41 & 37,37,37 & 37,37,36 & 37,37,36 & 37,37,37 & 37,37,37 & 37,37,37 & 37,37,37 &  \\
105,000 & 41,41,41 & 41,41,41 & 41,41,41 & 40,41,41 & 40,40,41 & 37,37,37 & 37,37,37 & 37,37,37 & 37,37,37 & 37,37,37 & 37,37,37 &  \\

110,000 & 41,41,41 & 41,41,41 & 40,41,41 & 40,41,41 & 40,41,41 & 37,41,41 & 37,37,41 & 37,37,37 & 37,37,37 & 37,37,37 & 37,37,37 &  \\
115,000 & 41,41,41 & 41,41,41 & 41,41,41 & 40,41,41 & 40,41,41 & 37,41,41 & 37,41,41 & 37,37,41 & 37,37,37 & 37,37,37 & 37,37,37 &  \\

120,000 & 41,41,41 & 41,41,41 & 41,41,41 & 40,41,41 & 40,41,41 & 37,41,41 & 37,41,41 & 37,37,41 & 37,37,37 & 37,37,37 & 37,37,37 &  \\
125,000 & 41,41,41 & 41,41,41 & 41,41,41 & 40,41,41 & 40,41,41 & 37,41,41 & 37,41,41 & 37,37,41 & 37,37,37 & 37,37,37 & 37,37,37 &  \\

130,000 & 41,41,41 & 41,41,41 & 40,41,41 & 40,41,41 & 40,41,41 & 37,41,41 & 37,41,41 & 37,37,41 & 37,37,37 & 37,37,37 & 37,37,37 &  \\
135,000 & 41,41,41 & 41,41,41 & 40,41,41 & 40,41,41 & 40,41,41 & 37,41,41 & 37,41,41 & 37,37,41 & 37,37,37 & 37,37,37 & 37,37,37 &  \\

140,000 & 41,41,41 & 41,41,41 & 40,41,41 & 40,41,41 & 40,41,41 & 37,41,41 & 37,41,41 & 37,41,41 & 37,37,41 & 37,37,37 & 37,37,37 &  \\
145,000 & 41,41,41 & 41,41,41 & 40,41,41 & 40,41,41 & 40,41,41 & 40,41,41 & 37,41,41 & 37,41,41 & 37,41,41 & 37,41,41 & 37,41,41 &  \\

150,000 & 41,41,41 & 41,41,41 & 40,41,41 & 40,41,41 & 40,41,41 & 40,41,41 & 37,41,41 & 37,41,41 & 37,41,41 & 37,41,41 & 37,41,41 &  \\
\enddata



\tablecomments{The first number in each tab is the optimal OPD for
$T_{\rm{eff}}=2400K$, the second number is for $T_{\rm{eff}}=2800K$ and the
third one is for $T_{\rm{eff}}=3100K$. OPD is in the unit of mm. OPD ranging
from 10mm to 41mm is considered in calculation.}


\end{deluxetable}


\begin{table}
\caption{Power Law Index $\chi$ as a function of Spectral Resolution
$R$ for DFDI and DE ($T_{\rm{eff}}=2400K$) \label{tab:PowerLawR}} \small
\begin{tabular}{c|c|ccc|ccc} \hline
\multicolumn{2}{c}{} & \multicolumn{3}{c}{DFDI} & \multicolumn{3}{c}{DE} \\
 \hline
$V \sin i$ & \multirow{2}{*}{$R$} & \multirow{2}{*}{5,000-20,000} & \multirow{2}{*}{20,000-50,000} & \multirow{2}{*}{50,000-150,000} & \multirow{2}{*}{5,000-20,000} & \multirow{2}{*}{20,000-50,000} & \multirow{2}{*}{50,000-150,000} \\

[$\rm{km\cdot s}^{-1}$] &  &  &  &  &  &  &  \\
\hline
 0 & \multirow{4}{*}{$\chi$} & 0.62 & 0.59 & 0.31 & 1.08 & 0.93 & 0.44 \\
 2 & & 0.63 & 0.56 & 0.27 & 1.07 & 0.89 & 0.38 \\
 5 & & 0.62 & 0.45 & 0.16 & 1.01 & 0.69 & 0.21 \\
 10 & & 0.58 & 0.28 & 0.09 & 0.87 & 0.39 & 0.10 \\
\hline
\end{tabular}
\end{table}

\begin{deluxetable}{ccc}

\tabletypesize{\footnotesize}
\tablewidth{355pt}

\tablecaption{Spectral Resolution and wavelength coverage on a given
detector}

\tablehead{\colhead{$R$} & \colhead{$\Delta\lambda$} & \colhead{$\lambda_{min}-\lambda_{max}$} \\
\colhead{} & \colhead{(nm)} & \colhead{(nm)} }

\startdata \label{tab:R_DeltaLambda}
25,000 & 480 & 800$-$1280 \\
30,000 & 400 & 800$-$1200 \\
40,000 & 300 & 850$-$1150 \\
50,000 & 240 & 880$-$1120 \\
60,000 & 200 & 900$-$1110 \\
70,000 & 170 & 910$-$1080 \\
80,000 & 150 & 920$-$1070 \\
\enddata

\end{deluxetable}

\clearpage

\begin{table}
\caption{$Q^{\prime\prime}$ comparison of DFDI and DE as a function of $V\sin i$ \label{tab:QPrimePrime}} \small
\begin{tabular}{c|c|cc|cc|c} \hline
\multicolumn{2}{c}{} & \multicolumn{2}{c}{DFDI} & \multicolumn{2}{c}{DE} &  \\
 \hline
 & $V\sin i$ [$\rm{km\cdot s}^{-1}$] & $R_{optimal}$ & $Q^{\prime\prime}_{\rm{DFDI}}$ &  $R_{optimal}$ & $Q^{\prime\prime}_{\rm{DE}}$ & $Q^{\prime\prime}_{\rm{DFDI}}$/$Q^{\prime\prime}_{\rm{DE}}$ \\
\hline
\multirow{4}{*}{$\alpha$=0.5} & 0 & 50,000 & 7502 & 110,000 & 6623 & 1.065 \\
 & 2 & 50,000 & 6806 & 75,000 & 6001 & 1.134 \\
 & 5 & 25,000 & 4979 & 50,000 & 4424 & 1.125 \\
  & 10 & 15,000 & 3394 & 25,000 & 2996 & 1.133 \\
\hline
 \multirow{4}{*}{$\alpha$=1.0} & 0 & 5,000 & 26384 & 30,000 & 8705 & 3.031 \\
 & 2 & 5,000 & 24297 & 25,000 & 8490 & 2.862 \\
 & 5 & 5,000 & 18929 & 15,000 & 7884 & 2.401 \\
 & 10 & 5,000 & 12877 & 10,000 & 7239 & 1.779 \\
\hline
\end{tabular}
\end{table}

\begin{deluxetable}{ccccccc}


\tabletypesize{\footnotesize}

\tablewidth{340pt}


\tablecaption{Predicted IRET Performance}

\tablehead{\colhead{Name} & \colhead{$m_J$} & \colhead{$T_{\rm{eff}}$} & \colhead{$V \sin{i}^a$} & \colhead{K} & \colhead{${\delta v_{rms,S}}^b$} & \colhead{${K_{\rm{HZ}}}^c$} \\
\colhead{} & \colhead{} & \colhead{(K)} & \colhead{($\rm{km\cdot s}^{-1}$)} &
\colhead{($\rm{m\cdot s}^{-1}$)} & \colhead{($\rm{m\cdot s}^{-1}$)} & \colhead{($\rm{m\cdot s}^{-1}$)} }

\startdata \label{tab:Detected_M_planet}
GJ 1214 b\footnotemark[1] & 9.75 & 3000 & 2 & 12 & 2.4 & 1.0 \\
GJ 176 b\footnotemark[2] & 6.46 & 3500 & 1 & 4.1 & 0.58 & 0.57 \\
GJ 179 b\footnotemark[3] & 7.81 & 3400 & 1 & 26 & 1.05 & 0.66 \\
GJ 436 b\footnotemark[4] & 6.9 & 3684$^d$ & 1 & 18.7 & 0.71 & 0.59 \\
HIP 57050 b\footnotemark[5] & 7.61 & 3190 & 1 & 38 & 0.91 & 0.68 \\
GJ 649 b\footnotemark[6] & 6.45 & 3700$^d$ & 1 & 12 & 0.58 & 0.54 \\

\enddata


\tablecomments{a: $V \sin{i}$ is assumed to be 1 $\rm{km\cdot s}^{-1}$ if otherwise
specified in references; b: the fundamental photon-limited RV
uncertainty; c: velocity semi-amplitude if there is a habitable
Earth-like planet locating at 0.05 AU from a host star; d: we assume
$T_{\rm{eff}}$ to be 3500K because we do not have synthetic stellar
spectrum with $T_{\rm{eff}}$ higher than 3500K.}

\tablerefs{1, ~\citet{Charbonneau2009}; 2, ~\citet{Forveille2009};
3, ~\citet{Howard2010}; 4, ~\citet{Butler2004}; 5,
~\citet{Haghighipour2010}; 6, ~\citet{Johnson2010}
}

\end{deluxetable}

\begin{deluxetable}{cc}


\tabletypesize{\footnotesize}

\tablewidth{320pt}


\tablecaption{Mid-Late M Dwarfs Available for IRET$^a$}


\tablehead{\colhead{$m_J$} & \colhead{Number of Targets$^b$} \\
\colhead{} & \colhead{} }

\startdata \label{tab:M_dwarf_targets}
J$\le$6 &  25 \\
J$\le$7 & 97 \\
J$\le$8 & 270 \\
J$\le$9 & 868 \\
J$\le$10 & 2458 \\
J$\le$11 & 6267 \\
J$\le$12 & 12775 \\
\enddata


\tablecomments{a: LSPM Catalog~\citep{Lepine2005}; b: we apply both
color cut (V-J$\ge$3) and $m_J$ cut.}


\end{deluxetable}

\end{document}